\documentclass[a4paper,fleqn,usenatbib]{mnras}

\usepackage{graphicx}	
\usepackage{amsmath}	
\usepackage{amssymb} 
\usepackage{multirow}

\title[\textit{AstroSat and NuSTAR observations of GRS 1758$-$258 and 1E 1740.7$-$2942}]{Broadband X-ray properties of black holes GRS 1758$-$258 and 1E 1740.7$-$2942: \textit{AstroSat} and \textit{NuSTAR} results}

\author[Bhuvana et al.]{
Bhuvana G. R.$^{1}$\thanks{E-mail: bhuvanahebbar@gmail.com,bhuvanagr.res-physics@dsu.edu.in},
Aneesha U.$^{2}$,
Radhika D.$^{1}$,
Vivek K. Agrawal$^{3}$,
Samir Mandal $^{4}$,
\newauthor
Tilak Katoch$^{5}$ and Anuj Nandi$^{3}$\\
$^{1}$Department of Physics, Dayananda Sagar University, Bengaluru, 560068\\
$^{2}$Indian Institute of Technology Guwahati, Guwahati, 781039, India \\
$^{3}$Space Astronomy Group, ISITE Campus, U R Rao Satellite Centre, Bengaluru, 560037\\
$^{4}$Department of Earth \& Space sciences, Indian Institute of Space science and Technology, Thiruvananthapuram, 695547\\
$^{5}$Department of Astronomy \& Astrophysics, Tata Institute of Fundamental Research, Mumbai, 400005 }

\date{Accepted XXX. Received YYY; in original form ZZZ}

\pubyear{2019}

\begin{document}
\label{firstpage}
\pagerange{\pageref{firstpage}--\pageref{lastpage}}
\maketitle

\begin{abstract}

We present the results on broadband X-ray properties of persistent black hole binaries GRS 1758$-$258 and 1E 1740.7$-$2942 using \textit{AstroSat}, \textit{NuSTAR} and \textit{Swift-XRT} observations carried out during 2016$-$2022. We perform spectral modeling of both sources after eliminating the contamination in their \textit{LAXPC} spectra from nearby X-ray sources. Preliminary spectral modelling using Comptonization and line emission ($\sim$ 6.4 keV) models suggest that GRS 1758$-$258 occupies both dim-soft state ($kT_{bb}=0.37\pm0.01$ keV, $\Gamma\sim5.9$, $L_{bol}=1\%$ of Eddington luminosity L$_{Edd}$) and hard state ($\Gamma=1.64-2.22$, $kT_{e}$=4$-$45 keV, $L_{bol}$=1$-$5\% L$_{Edd}$) that requires a multi-colour disc blackbody model ($kT_{in}=0.54\pm0.01$ keV) occasionally. 1E 1740.7$-$2942 instead is found only in hard state ($\Gamma$=1.67$-$2.32, $kT_{e}$=5$-$16 keV, $L_{bol}$=1$-$2\%  L$_{Edd}$). Reflection properties of both sources are studied by applying relativistic reflection model \textit{RELXILL}  to the broadband spectra. Our results from \textit{AstroSat} and \textit{NuSTAR} consistently unveiled the presence of a Comptonizing region along with an ionized reflection region (ionization parameter $log\xi$=2.7$-$3.8 and 2.7$-$4.7 erg cm s$^{-1}$ in GRS 1758$-$258 and 1E 1740.7$-$2942 respectively) in both sources. Reflection modeling revealed GRS 1758$-$258 to have a high metal abundance ($A_{fe}=3.9^{+0.4}_{-0.3}$ times solar metal abundance) and inclination angle ($i$) of $61\pm2^{\circ}$.  In case of 1E 1740.7$-$2942, $i$ is constrained to be $55\pm1^{\circ}$. Finally, we discuss the implication of our findings in the context of accretion dynamics by comparing our results with the previous studies.

\end{abstract}




\section{Introduction}

Stellar mass Black Holes (BHs) in our Galaxy are found to be in binary systems. They accrete matter from the companion and forms an accretion disc, which radiates X-rays. Such systems are known as X-ray binaries (XRBs), which are broadly classified either as Low Mass X-ray binaries (LMXBs) or as High Mass X-ray binaries (HMXBs) depending on the mass of the companion \citep{1995xrbi.nasa..126T,2016ApJS..222...15T,2021ApJ...921..131J}. Some of these XRBs exhibit continuous X-ray emission and hence can be classified as persistent binaries, while others undergo sudden outbursts lasting for months (\citealt{Chen97,2019MNRAS.487..928S,2020MNRAS.497.1197B} and references therein) and they are classified as transient sources. A few sources exhibit repeated transient behavior without periodicity and may not return to quiescent level for long periods of time \citep{2006ARA&A..44...49R,Tet2016}. Some transient LMXBs are observed to show consistent brightness following their outburst for decades and also exhibit aperiodic variability in their light curves \citep{C-T94,2019MNRAS.487..928S,2020MNRAS.499.5891S,2021MNRAS.507.2602K,2022MNRAS.510.3019A}. A catalog of all these sources and their salient characteristics are discussed by \cite{2016A&A...587A..61C,Tet2016}.

The black hole binaries exhibit different spectral and temporal variabilities by occupying low/hard, intermediate, or high/soft spectral states \citep{2001ApJS..132..377H,2005Ap&SS.300..107H,2006ARA&A..44...49R,Nandi2012,RN2014,2018Ap&SS.363...90N,2019MNRAS.487..928S,2021MNRAS.502.1334S}. The softer region of the energy spectrum is generally characterized by a multi-colored black body emission from the standard thin accretion disc \citep{1973A&A....24..337S}, while the high energy part is formed from the Comptonization of soft photons from a hot corona that exists around the BH (\citealt{1980A&A....86..121S,1995ApJ...455..623C,2015ApJ...807..108I} and references therein). The spectrum often shows additional re-emission features formed by the radiation reflected off the disc. Such reflection signature is found in the form of Fe-emission line at $\sim6.4$ keV \citep{1989MNRAS.238..729F} and Compton hump $\sim$ $15-30$ keV \citep{2016AN....337..375F} in the energy spectrum. Persistent sources occupy either a low/hard or high/soft state for long periods of time. These sources are usually HMXBs that accrete by means of stellar wind \citep{1995Obs...115..343K,Tet2016} or very rarely by Roche-lobe overflow \citep{2014ApJ...794..154O}. They also exhibit variability in their light curves. The power density spectra (PDS) obtained for such systems generally has low frequency breaks \citep{1990A&A...227L..33B} and Quasi-periodic Oscillation (QPO) features \citep{1989ARA&A..27..517V}. These BHs are observed to show spectral state changes on timescales of weeks to months (e.g.: Cyg X-1; \citealt{Geir99,Zdz02,2021MNRAS.507.2602K}) as well as seconds to minutes (e.g.: GRS 1915+105; \citealt{2000A&A...355..271B,2000ApJ...544..443R,2022MNRAS.510.3019A,2022MNRAS.512.2508M}). LMC X$-$1 is a persistent source that has been observed to remain steadily in soft state \citep{1989PASJ...41..519E,1999AJ....117.1292S,2021MNRAS.501.5457B,2022AdSpR..69..483B}, while state transitions have been reported in LMC X-3 (\citealt{2000ApJ...542L.127B, 2001MNRAS.320..327W, 2022AdSpR..69..483B}). Some of these persistent sources have bright lobes in relativistic jets that emit in radio and such sources are termed as `microquasars' \citep{Mir94,1994ApJ...430..829H}. 

The black hole binary GRS 1758$-$258 is one of the brightest persistent X-ray sources, which was discovered by {\it GRANAT} in 1990 \citep{Sun1991}. It was recognized as a microquasar based on the detection of its radio counterpart \citep{Rod1992,2017NatCo...8.1757M}. Distance to this source is estimated to be $8\pm1$ kpc by considering Jet-ISM interaction  \citep{2020MNRAS.497.3504T}. In contrast to a typical persistent source, GRS 1758$-$258 is a LMXB that accretes matter via Roche lobe overflow from its companion star \citep{2014ApJ...797L...1L}. Based on the long term observations performed by {\it RXTE}, the hydrogen column density ($N_{H}$) along the source line of sight is found to be $1.5 \pm 0.1 \times$ 10$^{22}$ atoms cm$^{-2}$ \citep{Mere97}. Such a high Galactic $N_{H}$ absorbs optical and infrared radiation from the system, because of which there has been no dynamical estimate of the BH mass yet. However, the luminosity-mass relation suggests that the system harbors a BH with mass $>8.6$ M$_{\odot}$ \citep{2001ApJ...563..301K}. GRS 1758$-$258 is mostly found in the hard state with the occasional transition into intermediate and dim-soft states \citep{2001AIPC..587...61G,2002ApJ...566..358M,2011MNRAS.415..410S}.
During the most commonly observed hard state, X-ray spectra consist of power-law component with photon index $\Gamma\sim1.5$  with an exponential cut-off at energies $>100$ keV \citep{Main99,2000ApJ...531..963L,2002A&A...388..293S,2011MNRAS.415..410S,2019MmSAI..90..100F}. Soft state and intermediate state observations are found to have a disc component present in the spectra where the photon index is greater than 2 \citep{2011MNRAS.415..410S,2020A&A...636A..51H}. During the soft state, the black body temperature is observed to be $\sim$ 0.4$-$0.5 keV \citep{2011MNRAS.415..410S}  which is low in comparison with the typical soft state disc temperature of $\sim1$ keV. Transition to this state leads to a drop in $L_{\rm bol}$ by a factor of 3 when compared to hard state luminosity, which is unusual, and hence this state is usually referred to as dim-soft state \citep{2001ATel...66....1S,2002ApJ...566..358M,Potts06}. Such peculiar spectral evolution of the source is not very
well understood and hence needs to be explored in detail.

The source 1E 1740.7$-$2942 is also a persistent black hole which is in a LMXB system that is considered as a microquasar \citep{1992Natur.358..215M}. It was discovered by the Einstein observatory \citep{HG1984} and is located in a crowded region i.e., closer to the Galactic center at a distance of $8\pm1$ kpc \citep{2020MNRAS.497.3504T}. It has high $N_{H}$ of $\sim$12$\times$10$^{22}$ atoms cm$^{-2}$ along the source line of sight due to the presence of a dense molecular cloud near the source \citep{1996ApJ...468..755S,1997ApJ...481..296Y,2002MNRAS.337..869G}. The source is typically found in its hard state, occasionally softening to intermediate and soft states \citep{Main99,2002ATel...94....1S,2005A&A...433..613D,Rey2010}. Long-term observations performed by {\it RXTE} suggest that the spectral photon index varies from 1.37 to 1.76 with the presence of an exponential cut-off at high energies \citep{Main99}. The energy spectrum is also reported to have the presence of Fe emission line and Compton hump due to the reflection of photons in the hard state along with cool disc and Comptonization components \citep{2005A&A...433..613D,Castro2014,2020MNRAS.493.2694S}. The soft state of the source is similar to that of GRS 1758$-$258, where the spectrum consists of a disc blackbody with temperature $\sim0.5$ keV, steep power-law component with index $>2$ and low bolometric flux \citep{2005A&A...433..613D}. Recent study of broadband X-ray observations from \textit{XMM Newton}, \textit{NuSTAR} and \textit{INTEGRAL} revealed that the system has a high inclination angle $i\geq50^{\circ}$, high spin, and BH mass of $\sim5$ M$_{\odot}$ \citep{2020MNRAS.493.2694S}.  Power spectral study using longer duration {\it RXTE} observations has shown the signature of QPOs with the PDS usually having a flat-top noise \citep{Main99}.

Broadband spectral coverage along with excellent timing capabilities sets out {\it AstroSat} (0.3$-$80 keV) \citep{Agrawal2001,JAA} and \textit{NuSTAR} (3$-$79 keV) \citep{2013ApJ...770..103H} as ideal observatories to study different characteristics of stellar-mass black hole binaries. \textit{Swift-XRT} (0.2$-$10 keV) has observed GRS 1758$-$258 simultaneously with \textit{NuSTAR}. Therefore, we use observations from these X-ray missions to carry out a spectral and temporal study of GRS 1758$-$258 and 1E 1740.7$-$2942. We explore the evolution of broadband energy spectrum during different Epochs of {\it AstroSat} and \textit{NuSTAR} observations within the period of 2016$-$2022. We look into the spectral properties of both sources along with spectral state transition of GRS 1758$-$258. We look for the evidence of reflection signature in their energy spectrum and find the relativistic reflection signature in both GRS 1758$-$258 and 1E 1740.7$-$2942. From the study of reflection spectra, we constrain the ionization parameter of the reflecting region as well as the inclination angle of both sources. 

This paper is organized as follows. We discuss the \textit{AstroSat} and \textit{NuSTAR} observations considered in this work and data reduction procedure in Section \ref{sec2}. In Section \ref{sec3}, we discuss the possible sources of contamination of the data and the procedure followed to eliminate it. Then in Section \ref{sec4}, we discuss the spectral analysis and different modeling procedures with results. Finally, in Section \ref{sec6}, we discuss and conclude the implications of the results obtained from this analysis.

\section{Observations and Data Reduction}
\label{sec2}
\begin{table*}
\centering
\caption{The table contains log of observations of GRS 1758$-$258 and 1E 1740.7$-$2942 by \textit{AstroSat} and \textit{NuSTAR} carried out during the period of 2016$-$2022. The exposure period of each observations is also given.}
\begin{tabular}{|c|c|c|c|c|c|c|c|c|}
\hline
Source                          & Date   & Mission                & Observation ID          & Epoch & \multicolumn{4}{c|}{\begin{tabular}[c]{@{}c@{}}Exposure\\ (ksec)\end{tabular}} \\ \hline
\multicolumn{5}{|c|}{}                                                                                & \textit{SXT}                     & \textit{LAXPC}                    & \textit{Swift-XRT}  & \textit{NuSTAR}                  \\ \hline
\multirow{6}{8em}{GRS 1758$-$258}   & 15-10-2016 & \multirow{4}{4em}{\textit{AstroSat}} & $A02\_077T01\_9000000732$ & AS1.1 & 58.6                    & 12.2                     & $-$  & $-$                       \\
                                & 28-07-2017 &                           & $A03\_053T01\_9000001410$ & AS1.2 & 43.4                    & 37.2                     & $-$      & $-$                   \\
                                & 20-09-2017 &                           & $A03\_053T01\_9000001542$ & AS1.3 & 48.2                    & 7.1                      & $-$        & $-$                 \\
                                & 08-04-2018 &                           & $G08\_070T01\_9000002028$ & AS1.4 & 7.5                     & 7.4                      & $-$          & $-$               \\
                                & 28-09-2018 & \textit{Swift-XRT }                   & 00088767001        & NU1.5 & $-$                       & $-$                        & 1.72 & $-$                      \\ 
                                &  & \textit{NuSTAR}                    & 30401030002             & & $-$                       & $-$                        & $-$ & 84.5                      \\ 
                                & 27-02-2022 &  \textit{Swift-XRT}                  & 00089378001    & NU1.6 & $-$                       & $-$                        & 1.69 & $-$   \\
                                 &  &  \textit{NuSTAR}                  &  90801302002             &  & $-$                       & $-$                        & $-$ &  23.9                      \\
                                \hline
\multirow{6}{8em}{1E 1740.7$-$2942} & 28-06-2016 & \multirow{5}{4em}{\textit{AstroSat}} & $G05\_158T01\_9000000630$ & AS2.1 & 88.6                    & 67.2                     & $-$                         \\
                                & 06-10-2016 &                           & $A02\_086T01\_9000000714$ & AS2.2 & 70.0                    & 67.2                     & $-$  & $-$                       \\
                                & 05-06-2018 &                           & G08\_045T01\_9000001920 & AS2.3 & 77.2                    & 73.9                     & $-$          & $-$               \\
                                & 11-05-2018 &                           & $G08\_045T01\_9000002092$ & AS2.4 & 68.6                    & 67.0                     & $-$        & $-$                 \\
                                & 12-05-2018 &                           & A04\_229T01\_9000002096 & AS2.5 & 94.5                    & 64.6                     & $-$           & $-$              \\
                                & 14-05-2021 & \textit{NuSTAR}                    & $90701317002$             & NU2.6 & $-$                       & $-$                        & $-$    & 43.9                  \\ \hline
\end{tabular}
\label{tab1}
\end{table*}

In this work, we make use of observations of the BH sources GRS 1758$-$258 and 1E 1740.7$-$2942 by \textit{Soft X-ray Telescope} ({\it SXT}) and \textit{Large Area X-ray Proportional Counter} ({\it LAXPC}) on-board {\it AstroSat}. We have also made use of two \textit{NuSTAR} observations of GRS 1758$-$258 and one observation of 1E 1740.7$-$2942. Along with these, we also consider two \textit{Swift-XRT} observations of GRS 1758$-$258 that are simultaneous with \textit{NuSTAR}. \textit{AstroSat} data are obtained through ISRO's archival web-page \footnote{\url{https://astrobrowse.issdc.gov.in/astro\_archive/archive/Home.jsp}} hosted at \texttt{ISSDC}.  We obtain the publicly available \textit{NuSTAR} and \textit{Swift-XRT} data observed within 2018$-$2022 from \texttt{HEASARC} database. We classify the considered observations of the sources into different Epochs as given in Table \ref{tab1}.

\subsection{\textit{NuSTAR} Data Reduction}

\textit{NuSTAR} \citep{2013ApJ...770..103H} observes the X-ray sources using two Focal Plane Modules, i.e., \textit{FPMA} and \textit{FPMB}, simultaneously in the energy band 3$-$79 keV. GRS 1758$-$258 has been observed during two occasions on 28-09-2018 and 27-02-2022, and 1E 1740.7$-$2942 is observed once on 14-05-2021 by \textit{NuSTAR}. We obtain these observations from the \texttt{HEASARC} database and process the data using \textit{NuSTAR} data analysis software \texttt{NuSTARDAS v 0.4.8} following standard reduction procedure\footnote{\url{https://heasarc.gsfc.nasa.gov/docs/nustar/analysis/}}. A cleaned event file is generated by using \texttt{nupipeline} module where screening of data is carried out by applying standard selection criteria. Latest \textit{NuSTAR} calibration database is used (CALDB version 20200813) to obtain the event file. From the cleaned event file, we obtain the source image using \texttt{XSELECT} tool and select a circular source region of radius 1.5$^{\prime}$ and a background region of 3$^{\prime}$ radius from a source free region. We extract source lightcurve and spectrum from this selected region using \texttt{nuproducts} module from both \textit{FPMA} and \textit{FPMB} (see \citealt{2018Ap&SS.363..189D,2022AdSpR..69..483B}). Response Matrix File (rmf) and Ancillary Response File (arf) required for the spectral fitting are also obtained using \texttt{nuproducts}. The background subtracted \textit{NuSTAR} spectra are regrouped to have a minimum signal-to-noise ratio of 10 in each bin.

\subsection{\textit{Swift-XRT} Data Reduction}

The \textit{Swift X-ray Telescope (XRT)} has observed GRS 1758$-$258 in 0.3$-$10 keV energy band, contemporaneously with the two \textit{NuSTAR} observations listed in Table \ref{tab1}. Both of these \textit{XRT} observations are carried out in Windowed Timing (WT) mode. We make use of standard data reduction pipeline associated with \texttt{HEASoft v6.30.1} to reduce the data. Since the source is not bright enough to cause pileup, we extracted the source spectrum from a circular region of radius 47${\arcsec}$ (20 pixel: 1 pixel=2.36${\arcsec}$). The background spectrum is extracted from a circular region of same radius in a source free region. We make use of these observations along with \textit{NuSTAR} to have wide-band coverage. The \textit{XRT} spectra are set to have minimal signal-to-noise ratio of 10 per bin after subtracting the background.

\subsection{\textit{SXT} Data Reduction}

The \textit{SXT} on-board \textit{AstroSat} observes the sources in 0.3$-$8.0 keV energy band. Both GRS 1758$-$258 and 1E 1740.7$-$2942 have been observed by {\it SXT} in the Photon Counting (PC) mode. We make use of the cleaned event file of the Level-2 data. We create a merged cleaned event file taking all the individual orbit files into account, using the event merger python routine\footnote{\url{https://www.tifr.res.in/~astrosat_sxt/dataanalysis.html}} based on {\it Julia v 1.1.1}. We select the single-pixel events by applying a grade $0$ filter to the event file to avoid the optical data leakage and to increase the signal-to-noise ratio \citep{2021JApA...42...77S}. From this merged and filtered event file, we generate the source image with the help of \texttt{XSELECT}. We select a circular region of 10$\arcmin$ around the source coordinates from which source spectra and lightcurves are extracted (see \citealt{2019MNRAS.487..928S,2021MNRAS.508.2447B}). The maximum count rate of both sources is well within the pile-up limit\footnote{\url{https://www.tifr.res.in/~astrosat_sxt/pdf/AstroSat-Handbook-v1.10.pdf}} for PC mode of 40 counts s$^{-1}$. The light curve for each orbit is generated considering the minimum bin time of 2.4 sec from the source region. We then choose only those data segments where the observation is continuous, thereby excluding the duration of South Atlantic Anomaly (SAA) and Earth occultation. For both sources, we make use of the sky background spectrum and the response provided by the {\it SXT} team\footnote{\url{https://www.tifr.res.in/~astrosat_sxt/dataanalysis.html}}. Using the specific python-based tool of \texttt{sxtarfmodule}, we obtain the ARF for individual observations considering the selected source region. The background subtracted spectrum is re-binned to have a minimum signal-to-noise ratio of 5 in each bin.

\subsection{\textit{LAXPC} Data Reduction}

The {\it LAXPC} consists of three identical proportional counter detector units {\it LAXPC10, LAXPC20, and LAXPC30} covering the energy range 3$-$80 keV \citep{2016ApJ...833...27Y,JAA,2017ApJS..231...10A}. We obtain Level-1 data from the ISSDC data archive, and extract the Level-2 files of energy spectrum, background spectrum and response matrix file using the latest version of single routine \textit{LAXPC} software \texttt{LaxpcSoftversion3.4.3}  \footnote{\url{http://www.tifr.res.in/~astrosat\_laxpc/LaxpcSoft.html}} following \citealt{2017ApJS..231...10A,2019MNRAS.487..928S,2020MNRAS.497.1197B,2021MNRAS.501.6123K}. For all the observations of both sources, we have considered only {\it LAXPC20}, as it has a steady gain. For each observation, the source and background spectra and lightcurves are extracted from the top layer of the {\it LAXPC} unit, considering a single event to minimize the instrument effects beyond 30 keV (see also \citealt{2018MNRAS.477.5437A,2021MNRAS.501.6123K}). The background subtracted \textit{LAXPC} spectra are grouped to have a minimum signal-to-noise ratio of 5 in each bin.

Since both GRS 1758$-$258 and 1E 1740.7$-$2942 are located close to the crowded field of the Galactic center, the possibility of other X-ray sources incidentally appearing within the wide \textit{LAXPC} field of view (FOV) is very high. As this could contaminate the high energy \textit{LAXPC} observations, before proceeding into further analysis, observational data of both sources should be checked and filtered for any possible contamination. Thus, in the following section, we discuss the appropriate methodology to quantify and exclude such contamination in these two sources.

\section{Contamination from nearby X-ray sources}

\label{sec3}
\subsection{GRS 1758-258}
\label{sec3.1}
\begin{figure}
	\centering
	\includegraphics[width=8cm]{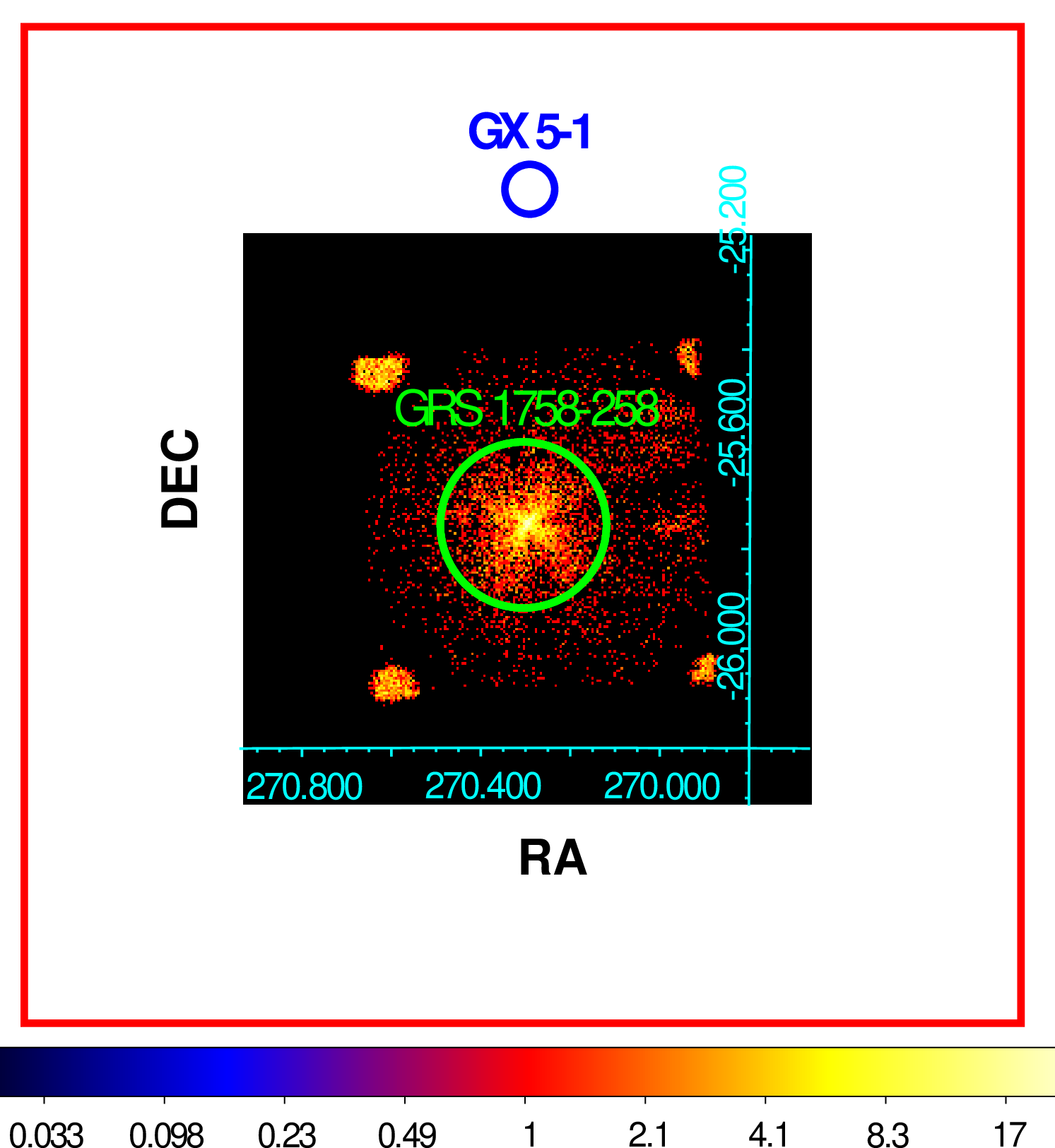}
	\caption{{\it SXT} image of GRS 1758$-$258 during the Epoch AS1.1 observation, where the source region of radius 10$^{\prime}$ is marked (green circle) along with the neighboring source GX 5$-$1 (blue circle). The red square indicates the {\it LAXPC} FOV ($2^{\circ}\times2^{\circ}$).}
\label{fig2}
\end{figure}

The source GRS 1758$-$258 is located in close proximity within an angular distance of $40^{\prime}$ from a persistent neutron star binary GX 5$-$1. Figure \ref{fig2} shows the {\it SXT} image of GRS 1758$-$258 along with GX 5$-$1 in the background during Epoch AS1.1. The green circle has a radius of $10^{\prime}$ from the central position corresponding to RA and Dec of GRS 1758$-$258, and it represents the \textit{SXT} region of selection. The nearby source GX 5$-$1 has been marked in a blue circle for reference. We have represented the {\it LAXPC} FOV of $2^{\circ}\times2^{\circ}$ (since its FWHM is $\sim1^{\circ}$) as a red square. It is evident that the source region of GX 5$-$1 is not close to the selected region in \textit{SXT} image of GRS 1758$-$258, implying there is no contamination of its {\it SXT} observations. However, we cannot rule out any possible contamination of its {\it LAXPC} observations since GX 5$-$1 lies within \textit{LAXPC} FOV. \textit{NuSTAR} and \textit{Swift-XRT} observations on the other hand are contamination-free because of their small FOV i.e., ($12.2{\arcmin}\times12.2{\arcmin}$)\footnote{\url{https://heasarc.gsfc.nasa.gov/docs/nustar/NuSTAR_observatory_guide-v1.0.pdf}} and ($23.6{\arcmin}\times23.6{\arcmin}$)\footnote{\url{https://swift.gsfc.nasa.gov/about_swift/xrt_desc.html}} respectively within which no other X-ray sources lie. At an angular separation of $40^{\prime}$, $30 \%$ of \textit{LAXPC} detector efficiency is expected \citep{2017ApJS..231...10A,2021MNRAS.501.6123K}. Therefore, for any further analysis of the \textit{LAXPC} spectrum, the contamination issue needs to be resolved.

In order to eliminate the contamination of \textit{LAXPC} spectrum of GRS 1758$-$258 by GX 5$-$1, we adapt the spectral subtraction method following \cite{2020MNRAS.497.1197B} and \cite{2021MNRAS.501.6123K}. In this method, contamination is quantified by simulating the spectrum of contaminating sources using an off-axis response file. The simulated spectrum thus produced is used as the background spectrum along with the primary background during the fitting of the source spectrum. For this purpose, we generate a fake spectrum of contaminating source i.e., GX 5$-$1 during different Epochs when GRS 1758$-$258 was observed following the procedure explained below.

\subsubsection{Simulation of Contamination from GX 5$-$1}

\label{sec3.1.1}

GX 5$-$1 is a Z-type neutron star binary that can occupy different spectral states in its various branches \citep{1991MNRAS.248..751V} i.e., Horizontal Branch (HB), Upper Normal Branch (UNB), Lower Normal Branch (LNB), and Flaring Branch (FB) of Hardness Intensity Diagram (HID) within timescales of a few kilo seconds \citep{2018ApJ...853..157H,2019RAA....19..114B}. In order to isolate the contamination by this source, it is necessary to simulate its spectrum, which entails understanding the spectral characteristic of the source. We simulate the \textit{LAXPC} spectrum of GX 5$-$1 for its several spectral states i.e., UNB, HB, LNB and FB by making use of spectral parameters obtained by \cite{2018ApJ...853..157H} using the model combination \textit{TBabs(diskbb+bbodyrad+cutoffpl)}. A \textit{LAXPC} off-axis response file for an off-axis angle of 40$^{\prime}$ provided by instrument team\footnote{\url{https://www.tifr.res.in/~astrosat_laxpc/LaxpcSoft_v1.0/}} is used for these simulations. The off-axis spectrum of GX 5-1 thus obtained, is then subtracted from the spectrum of target source by using it as background spectrum.

The variable source GX 5$-$1 could be in any of the above mentioned spectral states during our observations of GRS 1758$-$258. It is not possible to determine its exact spectral state during each of the \textit{AstroSat} Epochs as the source was not observed by any X-ray instruments during these periods. Hence, we determine the most appropriate state during each Epoch by including the individual GX 5$-$1 simulated spectrum of various states as background during the spectral fit and compare the \textit{LAXPC} spectral parameters with those obtained from uncontaminated \textit{SXT} in their overlapping energy range of 3$-$8 keV (see Appendix \ref{appena1}). From this, we find that the \textit{LAXPC} fit parameters are consistent with \textit{SXT} when background source is considered to be in UNB in all observations except Epoch AS1.1 (see Table \ref{taba1}), where the contamination is either over-corrected or under-corrected when different backgrounds are used. Therefore, for further analysis of Epoch AS1.1 observation, we ignore the \textit{LAXPC} and consider only the \textit{SXT} spectrum. For rest of the Epochs, we consider the contaminating source to be in UNB while carrying out the spectral analysis. 

\begin{figure}
\centering
\includegraphics[height=8.5cm,width=6cm,angle=-90]{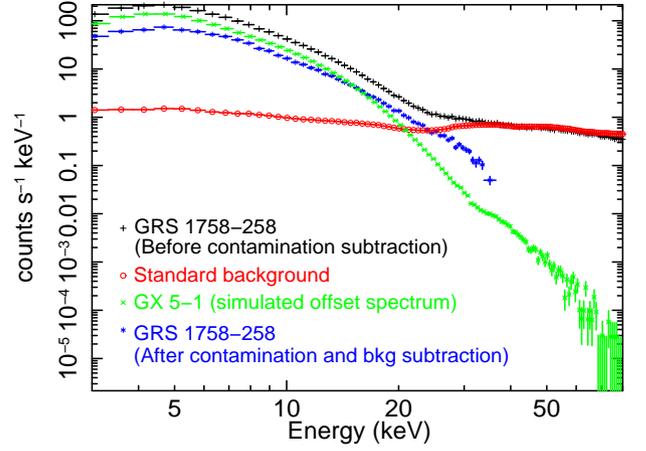}
\caption{Contaminated \textit{LAXPC} energy spectrum of GRS 1758$-$258 during Epoch AS1.3 is shown in black color. The standard \textit{LAXPC} background spectrum is plotted in red and the simulated off-axis \textit{LAXPC} energy spectrum of GX 5$-$1 is plotted in green colour. The energy spectrum of target source after the subtraction of both standard background and offset GX 5$-$1 spectrum is shown in blue color. See text for details.}
\label{fig5}
\end{figure}

The contamination free \textit{LAXPC} spectra are found to have significant data in high energy upto 30$-$40 keV in Epoch AS1.2, AS1.3 and AS1.4. We find the flux contribution of GX 5$-$1 in 3$-$30 keV to be $\sim65\%$, $\sim62\%$, and $\sim44\%$ of the total flux from GRS 1758$-$258 during Epoch AS1.2, AS1.3 and AS1.4 respectively. In Figure \ref{fig5}, we illustrate this by plotting the contaminated \textit{LAXPC} spectrum of GRS 1758$-$258 during Epoch AS1.3 (black) along with standard \textit{LAXPC} background spectrum for this observation (red), the simulated offset spectrum of GX 5$-$1 (green) as well as the source spectrum after the removal of standard background and contamination by GX 5$-$1 (blue).

\subsection{1E 1740.7-2942} 
\label{sec3.2}
\begin{figure}
	\centering
	\includegraphics[width=8cm]{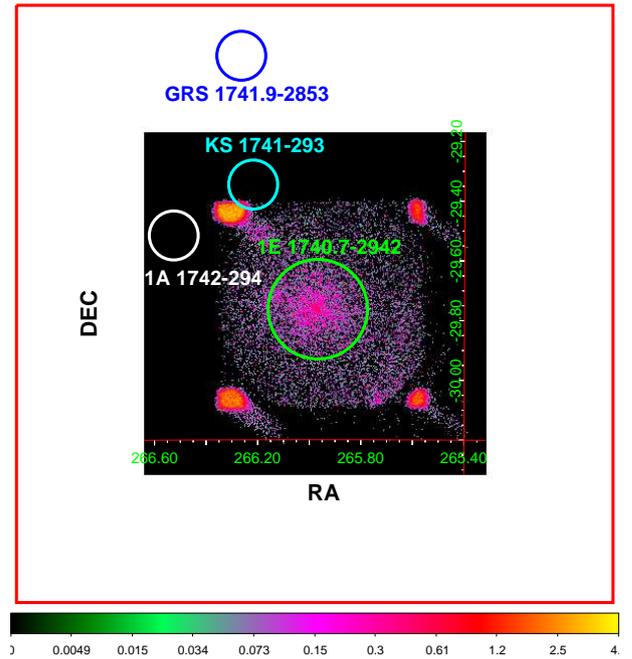}
	\caption{{\it SXT} image of 1E 1740.7$-$2942 during the Epoch AS2.2, where the source is marked using green circle along with the neighboring sources GRS 1741.9$-$2853, 1A 1742$-$294 and KS 1741$-$293 marked in blue, white and cyan circles respectively. The red box of dimension $2^{\circ}\times2^{\circ}$ indicates the FOV of {\it LAXPC}. See text for details.}
\label{fig}
\end{figure}

\begin{figure}
    \centering
    \includegraphics[width=8cm]{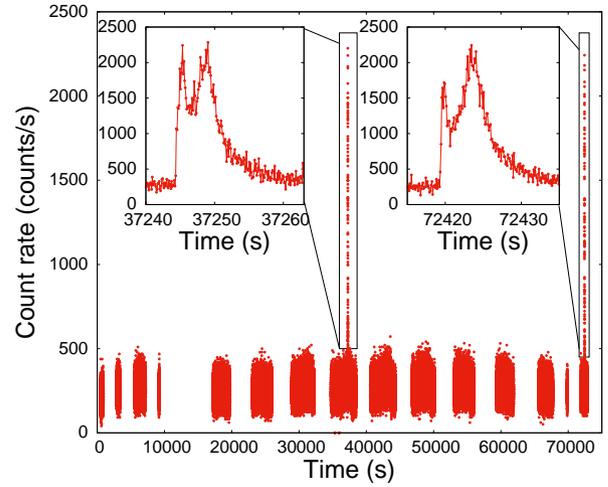}
    \caption{\textit{LAXPC} lightcurve of 1E 1740.7$-$2942 during Epoch AS2.2 plotted with a bin time of 0.1 sec. The inset panels show the expanded lightcurve of burst/flare seen during this Epoch.}
    \label{fig6}
\end{figure}

\begin{figure}
    \centering
    \includegraphics[height=8cm,keepaspectratio,angle=-90]{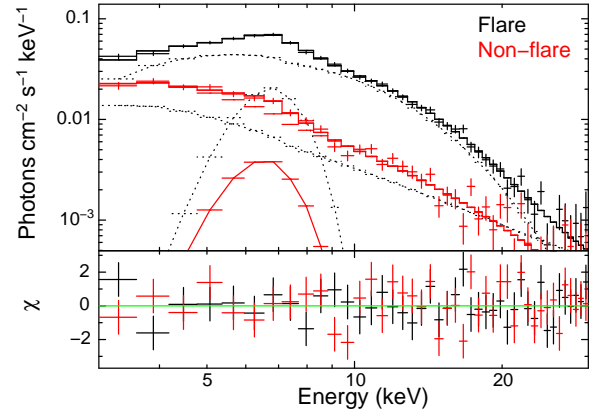}
    \caption{The unfolded energy spectrum of 1E 1740.7$-$2942 for the burst period of 20 sec (black) during Epoch AS2.2 that is modelled using \textit{TBabs(bbodyrad+nthComp+gauss)}. The energy spectrum (red) extracted from segment of 20 sec exposure period without flare is modelled using \textit{TBabs(nthComp+gauss)}. }
    \label{figflare}
\end{figure}

We also investigate the background X-ray sources in the vicinity of 1E 1740.7$-$2942, considering its proximity to the Galactic center. We find three neutron star (NS) binaries from the database of \textit{INTEGRAL} Galactic bulge survey\footnote{\url{http://hea.iki.rssi.ru/integral/nine-years-galactic-survey/index.php?key=GAL-Survey-GC}} namely, KS 1741$-$293, 1A 1742$-$294 and GRS 1741.9$-$2853 that are positioned at an angular distance of $\sim27^{\prime}$, $31^{\prime}$ and $50^{\prime}$ respectively from 1E 1740.7$-$2942 (see Appendix \ref{appen1}). Hence, they are identified as potential contaminators of 1E 1740.7$-$2942. In Figure \ref{fig}, we show the \textit{SXT} image of 1E 1740.7$-$2942 plotted along with KS 1741$-$293 (cyan circle), 1A 1742$-$294 (white circle) and GRS 1741.9$-$2853 (blue circle). It can be seen that these sources are located outside the region of selection of 1E 1740.7$-$2942 in \textit{SXT} FOV but well within the \textit{LAXPC} FOV, and hence we can expect the \textit{LAXPC} observations to be contaminated. Therefore, we inspect the \textit{LAXPC} lightcurve of 1E 1740.7$-$2942 and found it to have an occasional rapid increase in source count rate that lasts for a short period of $\sim20$ sec. Such burst-like feature in the \textit{LAXPC} lightcurve is observed to appear in every Epoch. However, no such traits are seen in the simultaneously observed \textit{SXT} lightcurves. Upon closer inspection of these bursts, it is observed to follow a fast rise and exponential decay burst profile with a duration of $\sim10-20$ sec. While some of these bursts show a single peak, a few others are found to have double peaks. This is illustrated in Figure \ref{fig6}, where we plot the source lightcurve during Epoch AS2.2 along with the expanded plots of burst profiles with double peaks in the inset. Such a profile is a characteristic signature of type-I thermonuclear burst from neutron stars (see \citealt{1993SSRv...62..223L} for details), which is not expected in a black hole system. Furthermore, we compare the spectral properties of flaring and non-flaring segments of \textit{LAXPC} data. We model the burst spectrum of 20 sec exposure period using \textit{TBabs(bbodyrad+nthComp+gauss)} in 3$-$30 keV energy range. Here, \textit{bbodyrad} models the spectrum as blackbody with single temperature, \textit{nthComp} models the Comptonization of the thermal radiation and \textit{gauss} is used to model the Fe-line emission present at $\sim6.4$ keV. While fitting, we tie up the blackbody temperature from \textit{bbodyrad} and seed photon temperature from \textit{nthComp}. This fitting resulted in a blackbody temperature of $\sim2.6$ keV and a spectral index of $\sim1.9$. We then generate a spectrum from a 20 sec data segment just prior to the onset of the burst during Epoch AS2.2 and carry out its modeling in 3$-$30 keV. We find that this spectrum can be fitted using \textit{TBabs(nthComp+gauss)} without the requirement of \textit{bbodyrad}. The seed photon temperature $kT_{bb}$ is fixed to 0.1 keV, as it could not be constrained. This fit resulted in a spectral index of $\sim2$ and Fe-line emission at $\sim6.4$ keV.  In Figure \ref{figflare}, we plot the unfolded energy spectra during a flare (black) and non-flare (red), which demonstrates an evident difference in the spectrum. Presence of a single temperature blackbody component in the burst spectrum that models the emission originating from the surface of the neutron star confirms the external origin of the flares. Even though the above mentioned nearby NS systems often undergo type-I burst with a similar burst profile \citep{2010ecsa.conf..222K,2013ATel.4840....1K}, we don't have enough evidence to prove which of these NS is the source of these bursts. Therefore, to avoid this contamination during the analysis of 1E 1740.7$-$2942, we extract the spectrum and lightcurve by excluding those segments with flaring signatures.  There is a possibility of the nearby NSs being active during non-flare periods also, which can cause contamination of \textit{LAXPC} spectrum of 1E 1740.7$-$2942. To identify the possibility of contamination, we study the spectral properties of each of the nearby neutron stars.


KS 1741$-$293 is a Very Faint X-ray Transient (VFXT) NS \citep{2021MNRAS.501.2790B} source located at an angular distance of $\sim27^{\prime}$ from 1E 1740.7$-$2942. The source is detectable in X-ray only when it undergoes an outburst during which the peak luminosity can reach $\sim10^{36}$ erg s$^{-1}$ \citep{2007MNRAS.380..615D,2011AstL...37..597C,2012A&A...545A..49D}. However, during the Epochs AS2.1$-$AS2.5, the source was in a quiescent/low-flux state, which is determined by referring to its \textit{JEM-X} lightcurve data. 

1A 1742$-$294 is a NS-XRB that is located at a distance of $\sim31^{\prime}$ from 1E 1740.7$-$2942 and is bright, persistent and exhibits type-I X-ray bursts. Because of its persistent nature and its brightness, the possibility of it causing contamination of  \textit{LAXPC} data of 1E 1740.7$-$2942 is high. In order to remove this undesired data, we simulate \textit{LAXPC} offset spectrum of 1A 1742$-$294 using $30^{\prime}$ offset response file. For the simulation, we use model combination \textit{TBabs(bremsstrahlung)} with $N_H= 6\times10^{22}$ cm$^{-2}$, $kT_{e}=10$ keV and flux (in 2$-$20 keV) $= 5.8 \times 10^{-10}$ erg s$^{-1}$  by referring \cite{1999ApJ...525..215S}. Contribution of flux from 1A 1742$-$294 to the total photons in 1E 1740.7$-$2942 spectrum in 3$-$30 keV are quantified to be 13.4\%, 10.9\%, 10.5\%, 11.5\% and 11.5\% during the Epochs AS2.1$-$AS2.5 respectively.

GRS 1741.9$-$2853 is a bursting NS-XRB located at a distance of $\sim50^{\prime}$ from 1E 1740.7$-$2942. It is a faint transient that emits luminosity of $10^{34} - 10^{36}$ erg s$^{-1}$  \citep{2005ATel..512....1W,2021ApJ...918....9P} during an outburst. We quantify the possible contamination from this source by using $50^{\prime}$ offset \textit{LAXPC} response file for the simulation, where we use the spectral parameters obtained by \cite{2021ApJ...918....9P} using the model combination \textit{TBabs(bbodyrad+nthComp+gauss)}. Contribution of GRS 1741.9$-$2853 is found to be 1.6\%, 1.4\%, 1.3\%, 1.5\%, and 1.5\%  to the total flux of 1E 1740.7$-$2942 (in 3$-$30 keV) during Epoch AS2.1$-$AS2.5 respectively. The simulated spectrum of GRS 1741.9$-$2853 is added to the 1A 1742$-$294 offset spectrum using \texttt{mathpha} tool which is then used as background along with the standard \textit{LAXPC} background. We see the total flux contribution from these NSs to be 15.1\%, 12.3\%, 12.9\%, 11.5\% and 12.2\% in 3$-$30 keV during the Epochs AS2.1$-$AS2.5 respectively. 

\begin{figure}
    \centering
    \includegraphics[width=6cm,height=8.5cm,angle=-90]{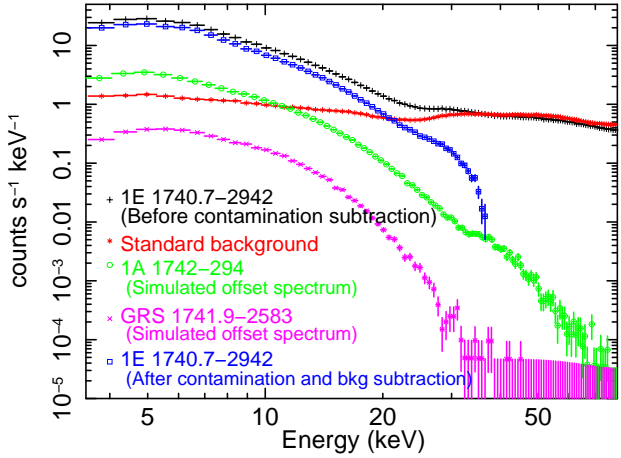}
    \caption{\textit{LAXPC} spectrum of 1E 1740.7$-$2942 belonging to Epoch AS2.1 observation before contamination subtraction is plotted in black. The standard \textit{LAXPC} background spectrum is shown in red and the simulated offset background spectrum of NS 1A 1742$-$294 and GRS 1741.9$-$2853 are shown in green and magenta respectively. The target source spectrum after the subtraction of background and contamination is plotted in blue colour. See text for details.}
    \label{fig3}
\end{figure}

To ensure the effective removal of contamination in \textit{LAXPC} spectra, we compare the \textit{LAXPC} and \textit{SXT} spectra of the target source in their overlapping energy band (i.e., 3$-$8 keV) in Appendix \ref{appena3}. We see that the flux and index in \textit{LAXPC} spectrum after the removal of contamination is close to that of \textit{SXT} (see Table \ref{tab1ea3}). This confirms the absence of overestimation/underestimation of contamination in \textit{LAXPC} spectrum of the target source. In Figure \ref{fig3}, we plot the contaminated source spectrum of 1E 1740.7$-$2942 (black) of Epoch AS2.1 along with standard \textit{LAXPC} background belonging to this observation (red), simulated offset spectra of both 1A 1742$-$294 (green) and GRS 1741.9$-$2853 (magenta) as well as the target source spectrum after the removal of contamination and standard background (blue). 

\section{Spectral Analysis and Results}
\label{sec4}

The spectral analysis is performed using \texttt{XSpec v 12.12.1} package of \texttt{HEAsoft v6.30}. A systematic error of 2\% is considered for all the \textit{AstroSat} spectral fits \citep{2019ApJ...871..152L,2019MNRAS.487..928S}. We incorporate gain corrections to all the \textit{SXT} fittings by using \texttt{gain fit} command in \texttt{XSpec} with fixed slope of 1 to correct for the instrumental residue at 1.8 and 2.2 keV. For those fitting where \texttt{gain fit} does not account for this residue, we use \textit{edge} models at 1.8 and 2.2 keV instead. All the \textit{AstroSat} spectral fits include \textit{gauss} at $\sim30$ keV to account for Xenon edge arising from the instrument \citep{2017ApJS..231...10A,2021JApA...42...32A}. We use \textit{TBabs} to account for the absorption of soft X-rays by neutral hydrogen in the interstellar medium. To estimate hydrogen column density $N_{H}$, abundance is set as per  \cite{2000ApJ...542..914W}. A cross-normalization constant is used between \textit{SXT} and \textit{LAXPC} which is fixed to unity in \textit{SXT} and is allowed to vary freely in \textit{LAXPC} data set whose value is found to be close to 1 in all the fits. Similarly, a cross-normalization is also used between \textit{XRT}, \textit{FPMA} and \textit{FPMB} for \textit{Swift-XRT $+$ NuSTAR} fitting by fixing \textit{FPMA} to unity. We find minimal cross normalization difference of $\sim1\%$ between \textit{FPMA} and \textit{FPMB} which is well within \textit{NuSTAR} accepted limit \citep{2015ApJS..220....8M}.

\begin{table*}
	\caption{Best-fit parameters of GRS 1758$-$258 obtained by carrying out preliminary spectral modeling of all the  \textit{AstroSat} and \textit{Swift-XRT+NuSTAR} spectra observed within 2016$-$2022. During the fit, the $N_{H}$ parameter is tied across the Epochs while rest of the parameters are allowed to vary independently. Here, $N_{dbb}$ is the \textit{diskbb} normalization, $lineE$ is the line energy and $\sigma$ is the gaussian line width. The errors of each parameters are quoted with 90$\%$ confidence.}
\begin{tabular}{cccccccc}
\hline
Model          & Parameter                                                           & Epoch AS1.1            & Epoch AS1.2                  & Epoch AS1.3            & Epoch AS1.4            & Epoch NU1.5           & Epoch NU1.6            \\ \hline
\texttt{TBabs}          & \begin{tabular}[c]{@{}c@{}}$N_{H}$ (atoms cm$^{-2}$) \end{tabular} & \multicolumn{6}{c}{$2.43\pm0.03\times10^{22}$}                                                                                                           \\
\texttt{diskbb}         & $kT_{in}$ (keV)                                                     & -                      & -                            & -                      & -                      & -                     & $0.54\pm0.01$ \\
               & $N_{dbb}$                                                       & -                      & -                            & -                      & -                      & -                     & $482^{+24}_{-107}$ \\
\texttt{nthComp}        & $kT_{bb}$ (keV)                                                     & $0.37\pm0.01$          & $0.1^{f}$                    & $0.1^{f}$              & $0.1^{f}$              & $0.1^{f}$             & $=kT_{in}$             \\
               & $\Gamma$                                                            & $5.90^{+0.19}_{-0.25}$ & $1.99\pm0.02$                & $2.22\pm0.02$          & $1.96\pm0.01$          & $1.64\pm0.01$         & $1.91\pm0.01$          \\
               & $kT_{e}$ (keV)                                                            & $15^{f}$               & $4.70^{+0.10}_{-0.11}$       & $5.42^{+0.08}_{-0.10}$ & $3.56^{+0.04}_{-0.05}$          & $44.8^{+1.8}_{-7.7}$ & $50^{f}$              \\
\texttt{gauss}          & $lineE$ (keV)                                                       &  $6.4^{f}$                    & $6.90^{+0.17}_{-0.16}$ & $6.94\pm0.11$ & $6.91\pm0.11$ & $6.00\pm0.19$         & $6.59\pm0.16$          \\
               & $\sigma$ (keV)                                                      &                      & $0.99^{*}$                      &               &               & $0.87\pm0.11$ & $1.42^{+0.11}_{-0.10}$ \\ \cline{2-8}
               & \begin{tabular}[c]{@{}c@{}}$F_{bol}^{\dagger} \times10^{-9}$\\ (erg cm$^{-2}$ s$^{-1}$)\end{tabular}                                                & 1.61                    & 3.81                         & 4.60                   & 8.04                  & 2.46                  & 1.58                   \\
               & \begin{tabular}[c]{@{}c@{}}$L_{bol}^{\ddag} \times10^{37}$\\ (erg s$^{-1}$)\end{tabular}                                                & 1.23                   & 2.92                         & 3.52                   & 6.16                   & 1.89                  & 1.21                   \\ \cline{2-8}
& $\chi^{2}$/dof &                                                                     & \multicolumn{5}{c}{5787.97/4796}                                                                                                                         \\ \hline
\end{tabular}
\label{tab2}
\raggedright
\footnotesize{$^{f}$ Frozen parameters} \\
\footnotesize{$^{*}$ Error couldn't be estimated} \\
\footnotesize{$^{\dagger}$ Bolometric flux derived in the energy range 0.3$-$100 keV} \\
\footnotesize{$^{\ddag}$ Bolometric luminosity calculated using $F_{bol}$ by assuming distance to the BH as 8 kpc \citep{2020MNRAS.497.3504T}}
\end{table*}

\begin{table*}
	\caption{Best-fit parameters of 1E 1740.7$-$2942 obtained by carrying out preliminary spectral modeling of all the  \textit{AstroSat} and \textit{NuSTAR} spectra observed within 2016$-$2022. The parameter $N_{H}$ is tied across the Epochs for this fit. The errors of each parameters are quoted with 90$\%$ confidence.}
\begin{tabular}{cccccccc}
\hline
Model                & Parameter                                                                                         & Epoch AS2.1            & Epoch AS2.2            & Epoch AS2.3            & Epoch AS2.4            & Epoch AS2.5                  & Epoch NU2.6            \\ \hline
\texttt{TBabs}                & $N_{H}$ (atoms cm$^{-2}$)                                                                         & \multicolumn{6}{c}{$10.8\pm0.1\times10^{22}$}                                                                                                                \\
\texttt{nthComp}              & $kT_{bb}$ (keV)                                                                                   & \multicolumn{6}{c}{$0.1^{f}$}                                                                                                                             \\
\multicolumn{1}{l}{} & $\Gamma$                                                                                          & $2.32\pm0.01$          & $1.81\pm0.20$ & $1.97^{+0.02}_{-0.01}$          & $1.99^{+0.01}_{-0.05}$          & $1.98\pm0.01$ & $1.67\pm0.01$          \\
                     & $kT_{e}$ (keV)                                                                                    & $8.78^{+0.18}_{-0.17}$ & $5.02^{+0.18}_{-0.13}$               & $4.95^{+0.20}_{-0.21}$ & $4.58^{+0.10}_{-0.02}$ & $50^{f}$                     & $15.8\pm0.8$ \\
\texttt{gauss}                & $lineE$ (keV)                                                                                     & $6.51^{+0.15}_{-0.18}$          & $6.40^{+0.14}_{-0.17}$          & $6.00\pm0.05$ & $6.00\pm0.02$ & $6.27^{+0.10}_{-0.12}$                & $6.00\pm0.11$          \\
                     & $\sigma$ (keV)                                                                                    & $1.22^{+0.15}_{-0.14}$                &               &                 &                 &                       & $1.08^{+0.14}_{-0.12}$     \\ \cline{2-8}
                     & \begin{tabular}[c]{@{}c@{}}$F_{bol}^{\dagger} \times10^{-9}$\\ (erg cm$^{-2}$ s$^{-1}$)\end{tabular} & 2.26                   & 1.45                   & 1.48                   & 1.49                   & 1.45                         & 0.65                   \\
                     & \begin{tabular}[c]{@{}c@{}}$L_{bol}^{\ddag} \times10^{37}$\\ (erg s$^{-1}$)\end{tabular}          & 1.73                   & 1.11                   & 1.13                   & 1.14                   & 1.11                        & 0.50                   \\ \cline{2-8}
&$\chi^{2}$/dof       &                                                                                                   & \multicolumn{5}{c}{2574.04/2239}                                                                                                                          \\ \hline
\end{tabular}\\
\label{tab3}
\raggedright
\footnotesize{$^{f}$ Frozen parameters} \\
\footnotesize{$^{\dagger}$ Bolometric flux derived in the energy range 0.3$-$100 keV} \\
\footnotesize{$^{\ddag}$ Bolometric luminosity calculated using $F_{bol}$ by assuming distance to the BH as 8 kpc \citep{2020MNRAS.497.3504T}} \\
\end{table*}
\subsection{Preliminary Spectral Analysis }
\label{sec4.1}

We model the simultaneous \textit{SXT} and \textit{LAXPC} energy spectra of GRS 1758$-$258 observed during four different Epochs and \textit{Swift-XRT $+$ NuSTAR} spectra during two Epochs (see Table \ref{tab1}). We use the standard \textit{LAXPC} background spectrum as we all as the simulated offset spectrum of GX 5$-$1 (see Section \ref{sec3.1}) as background file for the source spectra. As mentioned in Section \ref{sec3.1.1}, we consider only \textit{SXT} spectrum (0.6$-$8 keV) for the analysis of Epoch AS1.1 while rest of the fitting is carried out using combined \textit{SXT} and \textit{LAXPC} spectra in the energy range of 0.6$-$30 keV. High energy data bins are ignored as the source strength is weak and the background model is insensitive beyond 30 keV. Joint fitting of \textit{Swift-XRT} and \textit{NuSTAR} spectral fitting is carried out in the wideband energy range of 0.3$-$79 keV.

We begin the spectral modeling of GRS 1758$-$258 by adapting the basic thermal Comptonization model \textit{nthComp} \citep{1996MNRAS.283..193Z,1999MNRAS.309..561Z}. The \textit{nthComp} model describes the high energy continuum, which is originated from the process of up-scattering of seed photons coming from the disc. Here, the seed photons are modeled as multi-color disc blackbody. Photon index $\Gamma$, electron temperature $kT_{e}$ and seed photon temperature $kT_{bb}$ are the free parameters. Fitting this model to the \textit{AstroSat} and \textit{NuSTAR} energy spectra of GRS 1758$-$258 resulted in residual that shows a conspicuous structure of iron emission line at $\sim6.4$ keV (see also Appendix \ref{appenb}) in all of its \textit{LAXPC} and \textit{NuSTAR} spectra. We add a \textit{gauss} component to fit this emission feature. No such feature is observed in \textit{SXT} spectrum of Epoch AS1.1 due to limited spectral response beyond 6 keV. Hence, we attempt to derive an upper limit to the flux of possible presence of emission line by fixing line energy at 6.4 keV and line width ($\sigma$) at 1 keV. 
While the Comptonization model along with \textit{gauss} provides a good fit for most of the observation, it is seen that an additional thermal disc model is required to fit the lower energy only for Epoch NU1.6 spectrum (see Appendix \ref{appenb}). Thus, the preliminary spectral analysis of the source is carried out using \textit{(nthComp+gauss)} for Epoch AS1.1-NU1.5 and \textit{(diskbb+nthComp+gauss)} for Epoch NU1.6. Here, the $N_{H}$ parameters are tied across the Epochs while rest of the parameters are allowed to vary independently. We could not constrain source electron temperature during Epoch AS1.1 due to the unavailability of high energy data. Therefore, we freeze its value to $15$ keV as the fit remains unchanged beyond this value. The $kT_{bb}$ parameter in \textit{nthComp} is allowed to vary freely for Epoch AS1.1 whereas for Epoch AS1.2$-$NU1.5, when left free, $kT_{bb}$ hits the lower limit and hence we freeze its value to 0.1 keV in these Epochs. In Epoch NU1.6, $kT_{bb}$ is tied to the $kT_{in}$ of \textit{diskbb} which is left to vary freely. While fitting the \textit{gauss} model, line energy is limited to be within 6$-$7 keV. The $\sigma$ is left to vary independently across \textit{AstroSat} and \textit{NuSTAR} spectra but are tied within all the \textit{AstroSat} epochs. Leaving the $\sigma$ unconstrained, the \textit{AstroSat} spectral fits were resulting in a large value of $>1.5$ keV. Hence, we limit its value to be within 0.5$-$1.0 keV based on previous results by \textit{RXTE-PCA} observations (see \citealt{2020A&A...636A..51H}) since \textit{PCA} has a spectral resolution comparable to that of \textit{LAXPC}.
We list the parameters obtained from these fits during all six Epochs in Table \ref{tab2}. 

\begin{figure}
    \centering
    \includegraphics[height=8.5cm,keepaspectratio,angle=-90,trim={0 1.5cm 0 0},clip]{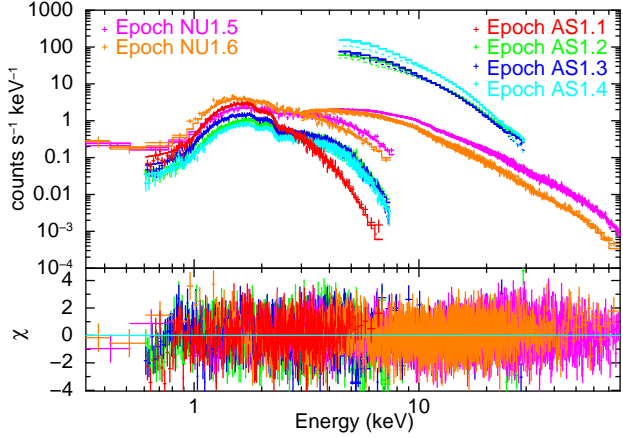}
    \caption{The energy spectra of GRS 1758$-$258 observed by \textit{AstroSat}, \textit{NuSTAR} and \textit{Swift-XRT} during six different Epochs modelled using an absorbed \textit{nthcomp} along with two \textit{gaussian} models to account for emission lines at $\sim$6.4 keV (Fe line) and $\sim30$ keV (instrument Xe line) and \textit{diskbb} for Epoch NU1.6 are plotted in top panel. The residual of this fit is plotted in the bottom panel. See text for details.}
    \label{fig7}
\end{figure}

From this preliminary fit, we find the $N_{H}$ value to be $2.43\pm0.03\times10^{22}$ atoms cm$^{-2}$. This value is slightly greater than the value reported by \citealt{Mere97} ($\sim1.5\times10^{22}$ atoms cm$^{-2}$) which is due to the consideration of distinct abundance i.e., from \citealt{1989GeCoA..53..197A} (see also \citealt{2022ApJ...936....3J}). The Fe-line energy is found to be within $6.0-6.94$ keV during these Epochs. The $\sigma$ in \textit{AstroSat} spectra results to upper limit of 1 keV whose 90\% confidence error couldn't be constrained, whereas in Epoch NU1.5 and NU1.6 resultant $\sigma$ is $0.87\pm0.11$ keV and 1.42$^{+0.11}_{-0.10}$ keV respectively. We obtain $kT_{bb}$ value of $0.37\pm0.01$ keV for Epoch AS1.1, where the spectrum is found to be very steep with $\Gamma=5.90^{+0.19}_{-0.25}$. The 90\% upper-limit to the Gaussian flux in this Epoch is $5.84\times10^{-12}$ erg cm$^{-1}$ s$^{-1}$. Unabsorbed source bolometric flux $F_{bol}$ in 0.3$-$100 keV during this Epoch is found to be $1.61\times10^{-9}$ erg cm$^{-2}$ s$^{-1}$. While thermal disc component is absent during Epoch AS1.2$-$NU1.5, it is present during Epoch NU1.6 where its temperature, $kT_{in}/kT_{bb}$ is found to be $0.54\pm0.01$ keV. We find relatively harder $\Gamma$ during Epochs AS1.2$-$NU1.6 whose value varies within $1.64-2.22$. The $kT_{e}$ value during these Epochs is found to be within $3.56-44.8$ keV. For Epoch NU1.6, we fix $kT_{e}$ to 50 keV as it could not be constrained when left free. $F_{bol}$ during AS1.2$-$NU1.5 is estimated to be within $2.46-8.04\times10^{-9}$ erg cm$^{-2}$ s$^{-1}$. During Epoch NU1.6, the overall flux is found to be $1.58\times10^{-9}$ erg cm$^{-2}$ s$^{-1}$. Total flux in this Epoch is dominated by Comptonized flux whose contribution is $\sim94\%$. In Figure \ref{fig7}, we plot the log data of the energy spectra during all six Epochs of GRS 1758$-$258. From this, it is evident that the energy spectrum during Epoch AS1.1 is much steeper than the other spectra. A cooler disc temperature and steeper $\Gamma$, along with a relatively lower flux value obtained during this Epoch suggests that the source is in a dim-soft state in this observation. Absence of disc component and presence of significant high energy contribution in the energy spectra of Epoch AS1.2$-$NU1.5 clearly indicate that the source was in a hard state during these periods. Although a disc component is seen during Epoch NU1.6, the spectrum is still hard, and moreover, the dominant Comptonized flux implies that this Epoch too belongs to a hard spectral state.

\begin{figure}
    \centering
    \includegraphics[height=8.5cm,keepaspectratio,angle=-90,trim={0 2.0cm 0 0},clip]{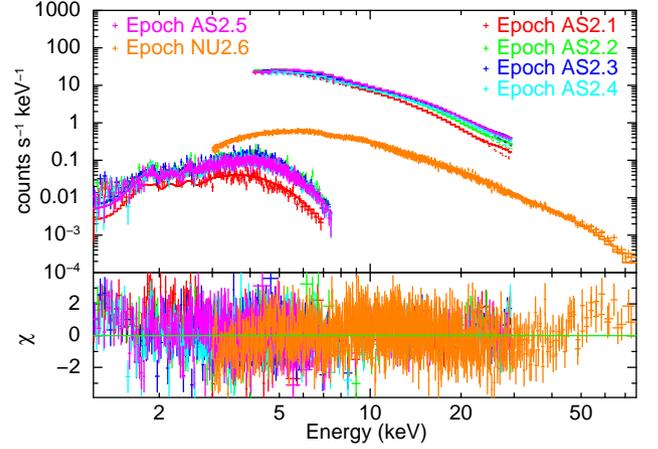}
    \caption{The energy spectra of 1E 1740.7$-$2942 observed by \textit{AstroSat} and \textit{NuSTAR} during all six Epochs modelled using absorbed \textit{nthcomp} and two \textit{gaussian} models to account for emission lines at $\sim$6.4 keV (Fe line) and $\sim30$ keV (instrument Xe line) are plotted in top panel. Residuals of these fits are plotted in bottom panel. See text for details.}
    \label{fig8}
\end{figure}

We perform spectral analysis of 1E 1740.7$-$2942 by fitting \textit{TBabs(nthComp+gauss)} model to the broadband {\it SXT} and {\it LAXPC} spectra belonging to Epoch AS2.1$-$AS2.5 and \textit{NuSTAR} spectrum belonging to Epoch NU2.6. As explained in Section \ref{sec3.2}, we have excluded all those segments in \textit{LAXPC} data that has burst signature while obtaining the energy spectrum. Background subtracted \textit{LAXPC} spectrum has data points till 30$-$40 keV in most of the observations. However, we carried out fitting of all the \textit{AstroSat} spectra in 0.8$-$30 keV ignoring the high energy bins for the same reason as mentioned for GRS 1758$-$258. \textit{NuSTAR} spectrum is modelled in the energy range 3$-$79 keV. Merged \textit{LAXPC} offset spectrum of 1A 1742$-$294 and GRS 1741.9$-$2853 is used as background along with standard \textit{LAXPC} background in all the \textit{AstroSat} spectral fits.  

We find that the disc component is absent during all Epochs of 1E 1740.7$-$2942 (F-test\footnote{\url{https://heasarc.gsfc.nasa.gov/xanadu/xspec/manual/node81.html}} probability of $\sim$0.34 for \textit{diskbb}). The parameters obtained by this spectral modeling during different Epochs are listed in Table \ref{tab3}. The parameter $N_{H}$ is found to be $10.8\pm0.1\times 10^{22}$ atoms cm$^{-2}$. We fix the $kT_{bb}$ to $0.1$ keV as it reaches a minimum value during all the Epochs. $\Gamma$ is found to vary in the range $1.67-2.32$. $kT_{e}$ couldn't be constrained during Epoch AS2.2 and AS2.5 therefore, we fix it to $50$ keV as fit remained unchanged beyond this value. During the rest of the Epochs, $kT_{e}$ is found to be within $4.58-15.8$ keV. In this source unlike GRS 1758$-$258, $\sigma$ could be constrained from \textit{AstroSat} spectra without limiting its value. We find $\sigma$ of $1.22^{+0.15}_{-0.14}$ keV and $1.08^{+0.14}_{-0.12}$ keV from \textit{AstroSat} and \textit{NuSTAR} spectra respectively. $F_{bol}$ during these observations are within the range of $0.65-2.26 \times 10^{-9}$ erg cm$^{-2}$ s$^{-1}$. The absence of a disc component in the spectrum and the flatter spectrum indicates that the source is in a hard state during all six observations. To illustrate the spectral fitting of \textit{AstroSat} and \textit{NuSTAR} observations, we plot the model fitted log data of spectra belonging to different Epochs in the top panel along with its residual in the bottom panel of  Figure \ref{fig8}, which resembles the typical hard state spectra of a BH-XRB.

\subsection{Relativistic Reflection Modeling}
\label{sec4.2}

In order to examine the reflection properties of GRS 1758$-$258 and 1E 1740.7$-$2942, we attempt to model the broadband \textit{AstroSat} and \textit{NuSTAR} spectra using several non-relativistic reflection models such as \textit{compPS} \citep{1996ApJ...470..249P}, \textit{pexriv} and \textit{pexrav}  \citep{1995MNRAS.273..837M}. Details of these modeling along with the results are discussed in Appendix \ref{appen}. The resultant non-zero values of the reflection parameter, i.e., $rel\_refl$ from these models implies the presence of a reflection component in the spectra of both sources. Further, the presence of broadened Fe-line in the source spectrum due to gravitational effects indicate the closeness of the inner disc region to the radius of the innermost circular orbit ($R_{ISCO}$) even during a hard state. As the relativistic effect has a significant influence on radiation from the region close to $R_{ISCO}$, in the present work, we focus on understanding the relativistic reflection properties of both sources. Thus, we model the source spectra obtained from \textit{AstroSat} and \textit{NuSTAR} using the relativistic reflection models belonging to \textit{relxill} (v.2.0) family \citep{2022MNRAS.514.3965D} that considers unconstrained coronal geometry. We fit the hard state spectra belonging to Epoch AS1.2 $-$ NU1.6 of GRS 1758$-$258 and Epoch AS2.1 $-$ NU2.6 of 1E 1740.7$-$2942 using self-consistent reflection model \textit{relxillCp} that includes both thermalized Compton emission from corona and the reprocessed emission from the disc. In this fit, we assume a single Comptonized emission by considering the equal emissivity indices ($q_{1}=q_{2}=3$) with the break radius fixed at standard value of 15 $R_{g}$ ($R_{g}=GM/c^{2}$). Inner disc radius $R_{in}$ is assumed to be extended all the way till $R_{ISCO}$ and outer disc radius $R_{out}$ is fixed at default value of $400$R$_{g}$. The parameters $N_{H}$, inclination angle ($i$) and disc iron abundance ($A_{fe}$) are tied across the Epochs since these parameters do not vary over time. The spectral index ($\Gamma$), electron temperature ($kT_{e}$), ionization parameter ($log \xi$) and reflection fraction ($R_{f}$) are allowed to vary freely. The electron density parameter $log N$ is frozen at the standard value of 15 as keeping it free did not yield any higher value. Along with \textit{relxillCp}, we use \textit{diskbb} for Epoch NU1.6 observation to account for thermal disc component. A \textit{gauss} model is used in all the \textit{AstroSat} observations to account for Xe emission at $\sim$ 30 keV. Along with these models, an additional \textit{gauss} is fitted at $\sim6.4$ keV in those observations where \textit{relxillCp} alone doesn't account for the complete Fe emission.

\begin{figure}
    \centering
    \includegraphics[angle=-90,width=8cm]{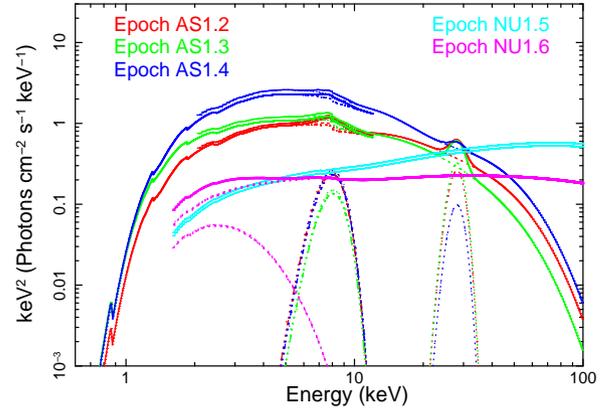}
    \caption{\textit{RelxillCp} model spectra used to fit the \textit{AstroSat} and \textit{NuSTAR} spectra of GRS 1758$-$258 belonging to different Epochs. \textit{Gauss} models fitted at $\sim6.4$ keV (Fe emission line) and $\sim30$ keV (instrument Xe line) are also marked. }
    \label{relxill}
\end{figure}

While fitting this model to GRS 1758$-$258, we assume the source to be maximally rotating in nature. Such an assumption is made based on the correlation between jet power in a BH and its spin \citep{1990agn..conf..161B}. Since GRS 1758$-$258 has a powerful jet (\citealt{2020MNRAS.497.3504T} and references therein), we consider it to be a maximally spinning BH. Therefore, fitting is carried out by fixing spin parameter ($a$) to $0.998$ while allowing $i$, $log \xi$, $R_{f}$ and $A_{fe}$ to vary freely. We list the model fitted parameters obtained during different Epochs in Table \ref{tab4}. It can be seen that for Epoch NU1.5 and NU1.6, additional \textit{gauss} is not required as \textit{relxillCp} alone accounts for the entire Fe emission line. From this reflection fitting, we find that the accretion disc in this system is partially ionized with the log $\xi$ value varying from 2.75$-$3.82 erg cm s$^{-1}$ during different Epochs. The $kT_{e}$ value is found to be $\sim3$ keV during Epoch AS1.2$-$AS1.4 and had to be frozen to 50 keV during Epoch NU1.5 and NU1.6. The $R_{f}$, which is the ratio of radiation emitted by corona towards the disc to the reflected intensity, is left to vary independently across the Epochs. Its value is found to be $>1$ i.e., 3.81$-$5.30 during Epoch AS1.2$-$AS1.4 which indicates presence of strong reflection whereas, $<1$ i.e., $0.56\pm0.10$ and $0.49^{+0.14}_{-0.06}$ during Epoch NU1.5 and NU1.6 respectively indicating a weaker reflection during these Epochs. The abundance of iron in accretion disc is found to be $3.89^{+0.37}_{-0.29}$ times the solar abundance. We also constrain the inclination angle ($i$) of this system to be $60.9\pm2^{\circ}$. We find that the reflection parameters obtained from \textit{AstroSat} and \textit{NuSTAR} spectral modelling are consistent with each other. In Figure \ref{relxill}, we show the \textit{TBabs(relxillCp+gauss)} model fit components present in \textit{AstroSat} and \textit{NuSTAR} spectra of GRS 1758$-$258 during different Epochs. \\
The additional \textit{gauss} component needed for three observations i.e., Epoch AS1.2, AS1.3 and AS1.4 could be a signature of distant reflection. To confirm this, we tried to replace \textit{gauss} with non-relativistic reflection model \textit{xillverCp}. However, two reflection models are too complicated for the data and as a result, most of the parameters couldn't be constrained. Therefore, we model the additional feature using \textit{gauss} model, whose line energy is found to be at $\sim6.9$ keV in all three observations. Similar to preliminary modelling, we restricted $\sigma$ to be within 0.5$-$1 keV which also resulted to its upper limit of 1 keV.

\begin{table*}
\caption{Spectral parameters obtained by fitting the relativistic reflection model to the hard state spectrum of GRS 1758$-$258 belonging to different Epochs. During the fit, the $N_{H}$, inclination angle ($i$) and iron abundance ($A_{fe}$) are tied across the Epochs. Here, the $i$ is expressed in degree and ionization parameter ($\xi$) in log erg cm s$^{-1}$. $A_{fe}$ is in the units of solar Fe abundance. Errors of all other parameters are obtained with 90$\%$ confidence.}
\label{tab4}
\begin{tabular}{ccccccc}
\hline
Spectral components & Parameter                            & Epoch AS1.2            & Epoch AS1.3             & Epoch AS1.4  & Epoch NU1.5       & Epoch NU1.6      \\ \hline
\texttt{TBabs}               & $N_{H}$   (atoms cm$^{-2}$) & \multicolumn{5}{c}{$1.89^{+0.04}_{-0.03}\times 10^{22}$}     \\ \hline
\texttt{relxillCp}           & $\Gamma$                             & $1.99^{+0.04}_{-0.05}$          & $2.28^{+0.04}_{-0.02}$  & $2.01^{+0.03}_{-0.02}$      & $1.63\pm0.01$      &   $1.90\pm0.01$  \\
                    & $log \xi$                            & $3.39\pm0.09$ & $3.82^{+0.31}_{-0.27}$  & $2.75^{+0.07}_{-0.06}$     & $3.70^{+0.15}_{-0.06}$  & $3.66^{+0.14}_{-0.23}$ \\
                    & $i$                                  & \multicolumn{5}{c}{$60.9\pm2$}  \\
                    & $A_{fe}$                             & \multicolumn{5}{c}{$3.89^{+0.37}_{-0.29}$ }            \\
                   
                    & $kT_{e}$  (keV)                           & $3.09^{+0.06}_{-0.05}$             & $3.60^{+0.11}_{-0.12}$              & $2.91^{+0.05}_{-0.04}$   & $50^{f}$ & $50^{f}$ \\
                    & $R_{f}$                              & $5.30\pm0.27$ &     $5.06^{+0.41}_{-0.38}$                    & $3.81^{+0.27}_{-0.40}$       &               $0.56\pm0.10$     &   $0.49^{+0.14}_{-0.06}$  \\
                    & norm $(\times10^{-3})^{\dagger}$                     & $5.14$                 & $16.5$           & $6.00$       & $3.63$ & $2.13$ \\ \hline 
                   \texttt{diskbb}           & $kT_{in}$ (keV)                            & $-$          & $-$  & $-$      & $-$               &  $0.68^{+0.06}_{-0.05}$    \\
                    & $N_{dbb}$                      & $-$                 & $-$  & $-$       & $-$                &   $60.9^{+12.2}_{-15.1}$  \\ \hline

                    & $\chi^{2}/dof$                       &  \multicolumn{5}{c}{$4660.94/4220$}         \\ \hline
\end{tabular}
\\
\raggedright
$^{f}$ Frozen parameter \\
$\dagger$ Errors are insignificant \\
\end{table*}

\begin{table*}
\caption{Fit parameters obtained by modeling the relativistic reflection model to the energy spectrum of 1E 1740.7$-$2942 belonging to Epoch AS2.1-NU2.6. During the fit, the $N_{H}$, inclination angle ($i$) and iron abundance ($A_{fe}$) are tied across the Epochs. Here, the $i$ is expressed in degree, ionization parameter ($\xi$) in log erg cm s$^{-1}$ and $A_{fe}$ is in the units of solar Fe abundance. Errors of all other parameters are obtained with 90$\%$ confidence.}
\label{tab5}
\begin{tabular}{cccllllc}
\hline
Spectral components                 & Parameter        & Epoch AS2.1            & Epoch AS2.2            & Epoch AS2.3           & Epoch AS2.4 & Epoch AS2.5             & Epoch NU2.6               \\ \hline
\texttt{TBabs}     & $N_{H}$ (atoms cm$^{-2}$)         & \multicolumn{6}{c}{$11.2\pm0.2\times10^{22}$} \\ \hline
\texttt{relxillCp} & $\Gamma$         & $2.18^{+0.05}_{-0.07}$ & $2.03^{+0.07}_{-0.05}$ & $1.75^{+0.03}_{-0.02}$ &     $1.74\pm0.05$        & $1.73\pm0.01$           & $1.61\pm0.01$             \\
                                    & $log \xi$        & $3.04^{+0.15}_{-0.06}$ & $3.50^{+0.19}_{-0.23}$         & $3.31\pm0.10$        &         $2.70^{+0.26}_{-0.09}$    & $3.26\pm0.11$  & $4.70^{+\ddag}_{-0.13}$  \\
                                    & $i$              & \multicolumn{6}{c}{$55.3^{+1.0}_{-1.1}$} \\
                                     & $kT_{e}$ (keV) &         $20^{f}$        &    $50^{f}$            &    $12^{f}$             &     $2.98^{+0.07}_{-0.06}$       &         $10^{f}$       &   $50^{f}$          \\
                                     & $R_{f}$ &         $2.65^{+1.15}_{-0.89}$        &    $3.11^{+0.10}_{-0.09}$            &    $2.65^{+0.91}_{-1.08}$             &     $4.13^{+0.18}_{-0.22}$       &         $1.92\pm0.34$       &   $1.13^{+0.89}_{-0.81}$          \\
                                      & norm $(\times10^{-3})^{\dagger}$                     & $5.03$                 &    $3.49$       & $1.64$       & $3.71$ & $1.10$ & 0.37 \\ \hline
                                    & $\chi^{2}/dof$   & \multicolumn{6}{c}{$2235.16/1986$}             \\ \hline
\end{tabular}
\\
\raggedright
$^{f}$ Frozen parameter \\
$\dagger$ Errors are insignificant \\
$^{\ddag}$ Parameter reaching hard limit \\
\end{table*}
\begin{figure}
    \centering
    \includegraphics[angle=-90,width=8cm]{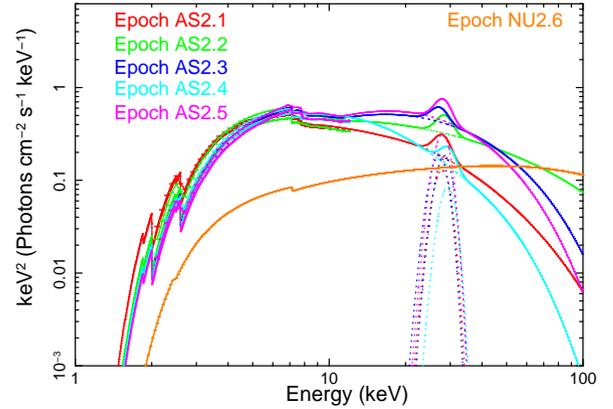}
    \caption{\textit{RelxillCp} model spectra used to fit the \textit{AstroSat} and \textit{NuSTAR} spectra of 1E 1740.7$-$2942 belonging to different Epochs. \textit{Gauss} model fitted at $\sim30$ keV (instrument Xe line) is also marked.}
    \label{fig10}
\end{figure}

We carry out the reflection modeling of \textit{AstroSat} and \textit{NuSTAR} energy spectra of 1E 1740.7$-$2942 in a similar way, using \textit{TBabs(relxillCp)}. In these observations, \textit{relxillCp} alone describes the entire spectrum without the need for an additional \textit{gauss}. In order to simplify the fitting, we fix the value of $A_{fe}$ at twice the solar abundance in all fits based on results by \cite{Rey2010}.
The parameters obtained from these fits are listed in Table \ref{tab5}. We assume the source to be maximally spinning BH \citep{2020MNRAS.493.2694S}. This allowed us to constrain the inclination angle of the system to be $55.3^{+1.0\circ}_{-1.1}$.The $log \xi$ from \textit{relxillCp} is found to be in the range 2.7$-$4.7  erg cm s$^{-1}$ and $R_{f}$ to be 1.13$-$4.13 during different Epochs. During most of these observations, $kT_{e}$ had to be frozen as we couldn't constrain its values. In Figure \ref{fig10}, we plot the total spectrum along with individual components fitted to \textit{AstroSat} and \textit{NuSTAR} spectra during different Epochs.

\section{Discussions and Conclusions}
\label{sec6}

In this paper, we have presented the spectral analysis of the persistent BH-XRBs GRS 1758$-$258 and 1E 1740.7$-$2942 using the broadband observations carried out by \textit{AstroSat}, \textit{Swift-XRT} and \textit{NuSTAR} during the period of 2016$-$2022. Here, we mainly discuss the source properties in different spectral states using \textit{nthComp+gauss} models and its reflection properties using \textit{relxillCp} model.

We find a bright NS source GX 5$-$1 located within \textit{LAXPC} FOV of GRS 1758$-$258 (see Figure \ref{fig2}) which could contaminate its \textit{LAXPC} spectra. Therefore, before proceeding with the spectral modeling, we eliminate the possible contamination of GRS 1758$-$258 spectra as discussed in Section \ref{sec3.1}. Preliminary spectral analysis of the uncontaminated data reveals the source to be in a dim-soft state during Epoch AS1.1 and a hard state during the rest of the Epochs. The hard state spectra are characterized by a Comptonization component along with Fe-line emission during Epoch AS1.2$-$NU1.5 with an additional disc component in Epoch NU1.6 (see Section \ref{sec4.1}). Hard state spectrum is also found to have a relativistic reflection component (see  Section \ref{sec4.2}). The abnormally high $\Gamma$ ($\sim5.9$), low inner disc photon temperature ($\sim0.37$ keV) and low flux ($F_{bol}=1.61\times10^{-9}$ erg cm$^{-2}$ s$^{-1}$) in its dim-soft state during Epoch AS1.1 is consistent with the previously observed dim-soft state of the source \citep{2002ApJ...566..358M,Potts06}. Such a soft state is significantly different from the typical soft state of Galactic BHBs, which usually has a disc temperature of $\sim1$ keV with a substantially higher flux than that in hard state (e.g., Cyg X-1; \citealt{2002ApJ...578..357Z,2021MNRAS.507.2602K}, IGR J1709$-$3624; \citealt{2018Ap&SS.363..189D}, LMC X-3; \citealt{2022AdSpR..69..483B}). Presence of dim-soft state has been interpreted by \cite{2001ApJ...554L..41S,2002ApJ...569..362S} as source having a distinct mechanism of spectral state change when compared to typical BHs like Cyg X-1. State change in GRS 1758$-$258 is understood to be due to a change in corona rather than a change in the accretion disc. Such scenario is explained by \cite{2001ApJ...554L..41S,2002ApJ...569..362S} based on Two-Component Accretion Flow (TCAF) model \citep{1995ApJ...455..623C}, where the BH source has independent accretion flow, i.e., Keplerian (thin disc) and sub-Keplerian (hot halo) disc which accretes the matter at different timescales. The change in the accretion rate affects the sub-Keplerian disc immediately, and the Keplerian disc catches up after weeks which forms a temporary soft state, i.e., a sudden drop in mass accretion is resulting in a dim-soft state. In addition to this argument, \citealt{Potts06} has discussed that the occurrence of a dim-soft state could be related to the orientation of the disc, i.e., dim-soft state is observed if the system orientation is close to edge-on which allows only a small fraction of flux from disc to reach the observer. The relatively lower value of inclination angle found in this study (see section \ref{sec4.2}) rules out this possibility.

During Epoch AS1.2$-$NU1.5, the hard state spectrum of GRS 1758$-$258 consists of only Comptonization component with cooler seed photon temperature ($<0.1$ keV) and a flatter spectrum ($\Gamma=1.6-2.2$). Hard spectra with low seed photon temperature attained in this source are consistent with the typical hard state behavior of Galactic BHs \citep{2004PThPS.155...99Z,2006ARA&A..44...49R,2021MNRAS.508.2447B,2022AdSpR..69..483B,2022MNRAS.510.3019A}. Hard state occurrence during Epoch NU1.6 is different from rest of the hard states because of the presence of a thermal disc component. While seed photon temperature or disc temperature is $<0.1$ keV in the hard state of the former kind, it is found to be $\sim0.54$ keV in the latter. The L$_{\rm bol}$ of 0.009 L$_{\rm Edd}$ (considering the mass of BH to be $10$ M$_{\odot}$) during Epoch NU1.6 is slightly lesser than what is observed in the previous Epoch, i.e., 0.01 L$_{\rm Edd}$ and $\sim5$ times lesser than the brightest hard state (Epoch AS1.4) in which L$_{\rm bol}=0.05$ L$_{\rm Edd}$. Both detection and non-detection of thermal disc components during the hard state of this source are consistent with the previous findings \citep{Potts06,2011MNRAS.415..410S}. The disc temperature ($\sim0.54$ keV) that is observed during hard state is also consistent with previously observed disc temperature of hard state in this source. Mere presence of thermal disc unlike other hard state observations in this study does not allow us to classify it to be an intermediate state, as such classification is made only when the contribution of the disc component is significant \citep{2001AIPC..587...61G,2011MNRAS.415..410S}. However, in our observation, flux contribution from the disc is weak (i.e., $\sim5\%$) and hence is considered as hard state.  The drastic change in $kT_{e}$ (4$-$45 keV) during the Epochs AS1.2$-$NU1.5 can be associated with the variability seen within the hard states. During Epoch NU1.5, when the source is in `canonical' hard state, the source exhibits low luminosity and high $kT_{e}$ (see Table \ref{tab2}). Similarly, when emitted luminosity is the highest, i.e., during Epoch AS1.4, $kT_{e}$ is found to have attained the lowest value (see Table \ref{tab2}). Thus, variation in $kT_{e}$ can be attributed to the variation within hard state which is also observed in other typical BH-XRBs such as GX 339$-$4, v404 Cyg etc., \citep{2009ApJ...690L..97P,2020ApJ...889L..18Y}. The variation in luminosity within the hard state in this source is also reported by \cite{2011MNRAS.415..410S}, where a flux difference of a factor of 2 is present between the hard state observed during 2001$-$2003 and 2008$-$2009. In fact, the hard state luminosity during 2008$-$2009 was slightly lesser than the dim-soft state of 2001$-$2003. \cite{2011MNRAS.415..410S} interprets this as change in the spectral state not driven by mass accretion rate alone, which further supports the TCAF scenario. Such behavior is imitated during our observations as well, by exhibiting slightly lower flux during hard state of Epoch NU1.6 than during the dim-soft state of Epoch AS1.1 (see Table \ref{tab2}). 

Driven by the presence of Fe-line feature in the source spectrum during hard states, we explore the possibility of the presence of high energy relativistic reflection component in the spectrum of GRS 1758$-$258. On this account, we model the source spectra obtained from \textit{AstroSat} and \textit{NuSTAR} using \textit{relxillCp}, that fits the continuum and broadened Fe-emission line, along with an additional \textit{gauss} whenever required (see Section \ref{sec4.2}, Figure \ref{relxill}). Here, \textit{relxillCp} models the flux from a single Comptonizing region with an unconstrained geometry that is reflected off the disc. Results obtained by fitting this model combination to the \textit{AstroSat} and \textit{NuSTAR} spectral fits suggest that the disc is partially ionized by resulting in $log\xi=2.75-3.82$ erg cm s$^{-1}$ which is responsible for the strong Fe-line emission. The resultant positive value of $R_{f}=3.8-5.3$ (see Section \ref{sec4.2}) obtained from this fit indicates the light-bending effect on reflected radiation that deflects the rays back to the disc \citep{2014MNRAS.444L.100D,2016A&A...590A..76D}. The low value ($0.5-0.6$) of $R_{f}$ obtained during Epoch NU1.5 and NU1.6 indicates the presence of relatively weaker reflection during these Epochs. We constrain the inclination angle of the system to be $60.9\pm2^{\circ}$ (see Table \ref{tab4}). This value is in close agreement with the jet inclination angle of this source estimated by \cite{2020A&A...643A.150L}. We find a high metal abundance of the accretion disc, i.e., $3.89^{+0.37}_{-0.29}$ in this source relative to solar iron abundance. Thus, relativistic reflection modeling implies that GRS 1758$-$258 has a partially ionized region in the accretion disc with high iron abundance, closer to the compact object, reflecting off the corona's radiation. The presence of contamination in this source's high-energy X-ray spectrum made it challenging to identify the reflection factor in the previous studies. However, careful detection and removal of contamination in this work helped us to model and constrain its parameters. Moreover, the reflection parameters obtained are found to be consistent across \textit{AstroSat} and \textit{NuSTAR} observations which consolidates the detection.

1E 1740.7$-$2942, often known as the twin source of GRS 1758$-$258 because of their identical properties, is also located around the Galactic center and hence has a high possibility of contamination of its observational data by any nearby X-ray sources. Presence of burst profiles in \textit{LAXPC} lightcurve of the source indicates such contamination from a nearby neutron star binaries (Figure \ref{fig}, see also Figure \ref{fig6}). We identify and eliminate contamination from nearby NS sources 1A 1742$-$294 and GRS 1741.9$-$2853 during the non-flaring period and ignore the \textit{LAXPC} data segments that show flaring features. Modeling these \textit{AstroSat} spectra along with the recent observation done by \textit{NuSTAR} shows that the source is in hard state during all six Epochs (see Figure \ref{fig8}, Section \ref{sec4.1}). The hard state energy spectra consist of a Comptonization component with $\Gamma=1.7-2.3$ and $kT_{e}=5-16$ keV, a distinct Fe-emission line feature at $\sim6.4$ keV, along with a high energy reflection component. Our detection of Fe-line is consistent with the previous \textit{Suzaku} \citep{Rey2010} and \textit{XMM-Newton} \citep{Castro2014} observations. During these hard states, $L_{\rm bol}$ is found to be 0.01$-$0.02 L$_{\rm Edd}$ considering the mass of the BH to be 5 M$_{\odot}$ and distance to be 8 kpc. Previous studies of the source using \textit{Suzaku} \citep{Rey2010}, \textit{NuSTAR} \citep{2015ApJ...813L..21N}, \textit{XMM Newton} and \textit{INTEGRAL} \citep{Castro2014} have shown the presence of disc component in the hard state where the inner radius of the disc is positioned close to the compact object \citep{Rey2010,2020MNRAS.493.2694S}. However, in our study, we did not find disc component in any of the Epochs. Truncation of the disc at larger radii could not be the reason for the absence of the disc, as we observe distinct disc reflection features in the spectrum. Other possible explanation for the non-detection of the thermal disc component could be the obscuration of accretion disc by corona in hard state. \cite{2002ApJ...569..362S} also favors this scenario to explain the geometry of the BH system based on the viscous timescale of spectral state change. 

Previous studies of 1E 1740.7$-$2942 using \textit{NuSTAR} and \textit{Suzaku} have reported the presence of weak reflection feature \citep{Rey2010,2020MNRAS.493.2694S} in the energy spectrum. In our study, modeling the spectra from \textit{AstroSat} and \textit{NuSTAR} with a blurred reflection model reveals the presence of relativistic reflection signature in the source (see Section \ref{sec4.2}). Model demonstrates that the source accretion disc is ionized having $\xi=2.7-4.7$ log erg cm s$^{-1}$. By obtaining $R_{f}>1$,  the reflected radiation is found to be affected by the extreme gravity in the region close to the compact object analogous to GRS 1758$-$258.  Although we could not constrain the spin value from this modeling, we constrain the inclination angle of the disc to be $i=55\pm1^{\circ}$ (see Table \ref{tab5}). This is consistent with the results obtained by \cite{2020MNRAS.493.2694S} where, modeling the broadband spectra of 1E 1740.7$-$2942 resulted in high inclination ($>50^{\circ}$) and maximum spin. Further, obtained inclination angle is consistent with the jet inclination angle in this system estimated by \cite{2015A&A...584A.122L}.

We have also modelled the source spectra using different non-relativistic models to understand the basic properties of reflecting and Comptonizing regions (see Appendix \ref{appen}). From Comptonization model \textit{compPS} which also accounts for reflection, we find the Comptonizing region in GRS 1758$-$258 to have large optical depth of $\tau=2.99\pm0.02$ and has a temperature of $54.5^{+6.6}_{-0.7}$ keV. We also estimate the $\tau$ value by calculating the Compton y-parameter using the relationship between $\Gamma$ and the y-parameter (see \citealt{2019MNRAS.489..366K} for details). Using the $\Gamma$ and $kT_{e}$ obtained from \textit{compPS} fitting, we obtain the y-parameter for Epoch NU1.5. Using this y-parameter, we calculate $\tau$ using the relation mentioned by \cite{1980A&A....86..121S}. From this, we estimate $\tau$ to be $\sim2.4$, which is consistent with the observed value. Model \textit{pexriv} and \textit{pexrav} gives similar fits resulting in $rel\_refl$ of $0.19\pm0.03$ and $0.25\pm0.05$ respectively. By fitting reflection models to the energy spectrum of 1E 1740.7$-$2942, from \textit{pexriv} and \textit{pexrav}, we find the amount of reflection from the disc in this source to be within 0.6$-$0.9 which is consistent with the previous studies \citep{2014ApJ...780...63N,2020MNRAS.493.2694S}. The Comptonizing region is found to have optical depth of $0.45^{+0.33}_{-0.09}$ (see Table \ref{tabc2}). These parameter values are close to that obtained by \cite{2020MNRAS.493.2694S} by fitting same model to the spectra obtained from \textit{NuSTAR}, \textit{XMM-Newton} and \textit{INTEGRAL}. Thus, using non-relativistic reflection models, we compute the physical parameters of the source that could not be estimated from the primary spectral fits.

We also looked into the timing features of both sources using \textit{NuSTAR} observations. Power spectra constructed using lightcurve in 3$-$79 keV energy band for the frequency range of  0.001$-$1 Hz. PDS is found to have flat-top noise till $\sim0.4$ Hz, after which the power decays. Power spectrum modeled using \textit{Lorentzian} yields total RMS of 13.2\% in GRS 1758$-$258 and 10.8\% in 1E 1740.7$-$2942. Timing studies of these sources using any of the \textit{AstroSat} observations are not attempted because of the possible contamination of its \textit{LAXPC} lightcurve.

Our findings from this study can be summarized as follows: \\
We present a detailed analysis of persistent, Galactic BH-XRBs GRS 1758$-$258 and 1E 1740.7$-$2942 using the X-ray observations from \textit{AstroSat} and \textit{NuSTAR}. From the spectral analysis, we find the sources to have the following properties:
\begin{itemize}
	\item GRS 1758$-$258 is found to occupy two different spectral states i.e., dim-soft and hard states with $L_{\rm bol}$ of 1\% L$_{\rm Edd}$ in the former state and 1$-$5\% L$_{\rm Edd}$ in the latter. Thermal Comptonization model alone describes the dim-soft state spectrum of the source. The hard state spectrum is characterized by a Comptonization component and an excess line emission at $\sim6.4$ keV along with occasional appearance of thermal disc component.
	\item Relativistic reflection feature is found in the \textit{AstroSat} and \textit{NuSTAR} energy spectrum of GRS 1758$-$258. Fitting the reflection spectra using \textit{relxillCp} allowed us to constrain the ionization parameter of the reflecting region to be $2.75-3.82$ log erg cm s$^{-1}$ with an iron abundance of $3.9^{+0.4}_{-0.3}$ times solar iron abundance. It also revealed the disc inclination angle to be $61\pm2^{\circ}$.
	\item Hard state spectrum of 1E 1740.7$-$2942 is found to be similar to that of GRS 1758$-$258, which has a Comptonized component, Fe emission line, and reflection component. $L_{\rm bol}$ during these hard states is found to be in the range 1$-$2 \% L$_{\rm Edd}$. 
	\item 1E 1740.7$-$2942 is also found to have relativistic reflection features in the energy spectrum, and modeling the same suggests that the source has ionized accretion disc with $log \xi=2.7-4.7$ erg cm s$^{-1}$. Inclination angle of the system is revealed to be $55\pm1^{\circ}$.
\end{itemize}

\section*{Acknowledgment}

We acknowledge the anonymous reviewer for providing useful comments and suggestions regarding our work that helped to improve the quality of the manuscript. We acknowledge the financial support of Indian Space Research Organization (ISRO) under AstroSat archival data utilization program Sanction order No. DS-2B-13013(2)/13/2019-Sec.2. BGR thanks Dr. Baishali Garai, the PI of this project for providing constant support to carry out this research. This publication uses data from the AstroSat mission of the ISRO archived at the Indian Space Science Data Centre (ISSDC). This work also made use of data from the \textit{NuSTAR} mission by the National Aeronautics and Space Administration. This work has been performed utilizing the calibration databases and auxiliary analysis tools developed, maintained and distributed by \textit{AstroSat-SXT} team with members from various institutions in India and abroad. Also, this research made use of software provided by the High Energy Astrophysics Science Archive Research Center (HEASARC) and NASA’s Astrophysics Data System Bibliographic Services. This research also made use of \textit{JEM-X} lightcurve data provided by \textit{INTEGRAL} Galactic Bulge Monitoring program. We also acknowledge the copyright holders of Astronomy \& Astrophysics journal for providing permission to modify and reprint a published from Kuulkers et al. 2007. AU acknowledges the financial support from ISRO sponsored project (DS\_2B-13013(2)/5/2020-Sec.2). VKA, AN also thank GH, SAG, DD, PDMSA and Director, URSC for encouragement and continuous support to carry out this research. TK acknowledge support of the Department of Atomic Energy, Government of India, under project no. 12 R\&D$-$TFR$-$5.02$-$0200.

Facilities: \textit{AstroSat}, \textit{NuSTAR}, \textit{INTEGRAL}

\section*{Data Availability}
This paper uses the \textit{AstroSat} archival data available in ISSDC website \url{https://astrobrowse.issdc.gov.in/astro_archive/archive/Home.jsp}. \textit{JEM-X} data is available at \url{http://integral.esac.esa.int/BULGE/SOURCES/} and \textit{NuSTAR} at \url{https://heasarc.gsfc.nasa.gov/cgi-bin/W3Browse/w3browse.pl}. \\




\bibliographystyle{mnras}
\bibliography{mnras_template} 





\appendix 
\section{Contamination of \textit{LAXPC} spectrum}
\subsection{Determination of spectral state of GX 5$-$1}
\label{appena1}
The highly variable background source GX 5$-$1 could be in any of its different spectral states during the Epochs when GRS 1758$-$258 was observed by \textit{AstroSat} (see section \ref{sec3.1.1}). Since the determination of its exact spectral state is not possible, we identify the most appropriate state that reproduces the contamination in \textit{LAXPC} spectrum, by comparing the flux and spectral parameters obtained from \textit{SXT} and \textit{LAXPC} with different background in their overlapping energy band of 3$-$8 keV. For this, we fit the 3$-$8 keV \textit{SXT} and \textit{LAXPC} energy spectra using an absorbed \textit{powerlaw+gauss} model, considering different \textit{LAXPC} contamination spectrum i.e., UNB, HB, LNB and FB. In Table \ref{taba1}, we list the parameters obtained from these fits for different observations. It is evident that, the powerlaw index ($\Gamma$) of \textit{LAXPC} spectrum has remained more or less constant across different background states. However, the flux in \textit{LAXPC} spectrum changes significantly with different background. But, it is to be noted that the \textit{LAXPC} and \textit{SXT} fluxes are consistent with each other when background is considered to be in UNB during Epoch AS1.2, AS1.3 and AS1.4. However, during Epoch AS1.1, the \textit{LAXPC} flux is significantly lesser than \textit{SXT} when background spectrum is considered to be UNB, HB and LNB indicating over-estimation of level of contamination, whereas for FB, flux in \textit{LAXPC} is higher than \textit{SXT} implying under-estimation of contamination. 
\begin{table*}
\caption{Comparison of \textit{powerlaw+gauss} model fit parameters of GRS 1758$-$258 in 3$-$8 keV from \textit{SXT} and \textit{LAXPC}, considering \textit{LAXPC} background source GX 5$-$1 in different spectral states. The flux is in the unit of erg cm$^{-2}$ s$^{-1}$.}
\label{taba1}
\begin{tabular}{ccccccccccc}
\hline
Epoch & \multicolumn{5}{c}{$\Gamma$}                                                                                               & \multicolumn{5}{c}{Flux ($\times10^{-10}$)}                                                                      \\ \hline
      & SXT                    & \multicolumn{4}{c}{LAXPC}                                                                         & SXT                  & \multicolumn{4}{c}{LAXPC}                                                                 \\ \hline
      &                        & UNB                    & HB                     & LNB                    & FB                     &                      & UNB                  & HB                   & LNB                  & FB                   \\ \hline
AS1.1 & $5.62^{+0.19}_{-0.18}$ & $5.92^{+0.20}_{-0.23}$ & $5.71^{+0.30}_{-0.29}$ & $5.10\pm0.20$ & $4.91\pm+0.06$ & 16.8$\pm0.2$ & $10.5\pm0.1$ & $13.7\pm0.1$ & $14.3\pm0.1$ & $19.7\pm0.1$ \\
AS1.2 & $1.92\pm0.07$          & $1.99^{+0.30}_{-0.23}$ & $1.94\pm0.10$          & $1.79^{+0.20}_{-0.17}$ & $1.93^{+0.21}_{-0.22}$ & 17.0$\pm0.4$         & 16.8$^{+0.3}_{-0.2}$ & $23.5\pm0.1$         & $24.5\pm0.1$         & $66.7\pm0.3$         \\
AS1.3 & $2.14\pm0.07$          & $1.83^{+0.31}_{-0.24}$ & $1.99\pm0.07$          & $1.85^{+0.09}_{-0.08}$ & $1.95^{+0.09}_{-0.08}$ & 18.8$\pm0.4$         & 19.1$\pm0.4$         & $26.0\pm0.1$         & $26.7\pm0.1$         & $29.9\pm0.1$         \\
AS1.4 & $1.95\pm0.06$          & $2.24^{+0.31}_{-0.26}$ & $2.21\pm0.07$          & $2.13^{+0.09}_{-0.08}$ & $2.16^{+0.09}_{-0.08}$ & 40.0$^{+1.2}_{-0.9}$ & 41.9$^{+1.0}_{-0.9}$ & $49.4\pm0.1$         & $50.1\pm0.1$         & $53.4\pm0.1$         \\ \hline
\end{tabular}
\end{table*}

\subsection{Neutron stars nearby 1E 1740.7-2942}
\label{appen1}
\begin{figure*}
    \centering
    \includegraphics[width=10cm,trim={2cm 3cm 0 0},clip]{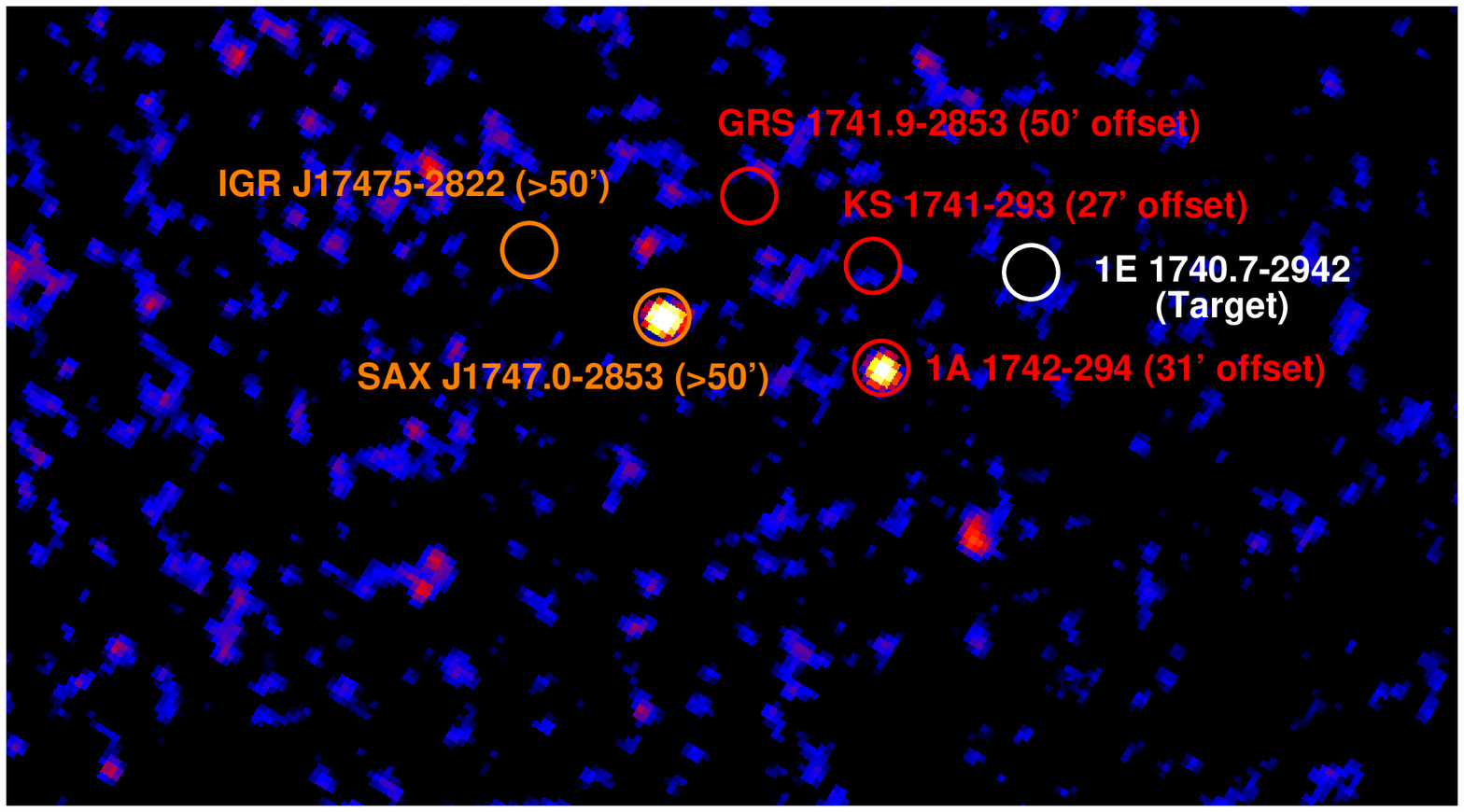}
    \caption{Image of Galactic centre plotted using \textit{JEM-X} (3$-$10 keV) observations. We have marked the target source 1E 1740.7$-$2942 in white along with contaminating sources KS 1741$-$293, 1A 1742$-$294 and GRS 1741.9$-$2853 in red. Farther sources, SAX J1747.0$-$2853 and IGR J17475$-$2822 are marked in orange colour. Credits: Kuulkers et al., A\&A, 466, 2, 595, 2007, reproduced with permission © ESO.}
    \label{appenA}
\end{figure*}

1E 1740.7$-$2942 is located in the midst of many outbursting neutron stars. In Figure \ref{appenA}, we  show the \textit{INTEGRAL/JEM-X} image of Galactic centre obtained by \cite{2007A&A...466..595K} with the marked nearby X-ray binaries. KS 1741$-$293 is the closest one that is at an angular distance of $\sim27^{\prime}$ from the target source. 1A 1742$-$294 is located at a distance of $\sim31^{\prime}$ and GRS 1741.9$-$2853 at $\sim50^{\prime}$. Other nearby X-ray sources IGR J17475$-$2822 and SAX J1747.0$-$2853 are at distance $>50^{\prime}$ and hence lie outside the FOV of \textit{LAXPC}. Therefore, there would be no contamination from these two sources.

\subsection{Comparison of \textit{SXT} and \textit{LAXPC} spectra of 1E 1740.7$-$2942 in their overlapping energy band}
\label{appena3}

In case of 1E 1740.7$-$2942, the background sources 1A 1742$-$294 and GRS 1741.9$-$2853 are considered to be in their brightest states. We did not account for the flux variability of these contaminating sources as the nature of 1A 1742$-$294 during non-burst emission is not very well studied and  GRS 1741.9$-$2853 is a very faint source whose variability doesn't affect the source flux significantly. Yet, we compare the \textit{LAXPC} spectral parameters with that of \textit{SXT} in 3$-$8 keV to check for overestimation/underestimation of level of contamination. We list the parameters obtained by fitting absorbed \textit{powerlaw+gauss} model to the individual \textit{SXT} and \textit{LAXPC} spectra during different Epochs in Table \ref{tab1ea3}. The power-law index as well as flux values obtained from \textit{LAXPC} spectrum remains consistent with that from \textit{SXT}, which demonstrates the effective removal of contamination in \textit{LAXPC} spectra of 1E 1740.7$-$2942.

\begin{table}
\caption{Parameters of 1E 1740.7$-$2942 obtained by fitting individual \textit{SXT} and \textit{LAXPC} spectrum during each Epochs using absorbed \textit{powerlaw+gauss} model in 3$-$8 keV. The flux is the unit of erg cm$^{-2}$ s$^{-1}$.}
\label{tab1ea3}
\begin{tabular}{ccccc}
\hline
Epoch             & \multicolumn{2}{c}{$\Gamma$}                    & \multicolumn{2}{c}{Flux ($\times10^{-10}$)}     \\ \hline
                  & SXT                    & LAXPC                  & SXT                    & LAXPC                  \\ \hline
AS2.1             & $2.4^{+0.24}_{-0.23}$  & $2.73^{+0.24}_{-0.29}$ & 7.76$^{+0.18}_{-0.14}$ & 7.54$^{+0.14}_{-0.08}$ \\
AS2.2             & $2.07^{+0.15}_{-0.15}$ & $2.27^{+0.10}_{-0.10}$ & 13.7$^{+2.0}_{-2.0}$   & 12.3$^{+1.6}_{-0.8}$   \\
AS2.3$^{\dagger}$ & $1.55^{+0.15}_{-0.15}$ & $1.96^{+0.37}_{-0.35}$ & 9.32$^{+0.36}_{-0.36}$ & 10.9$^{+2.2}_{-1.7}$   \\
AS2.4$^{\dagger}$ & $2.02^{+0.48}_{-0.49}$ & $2.09^{+0.07}_{-0.06}$ & 9.67$^{+0.27}_{-0.29}$ & 9.90$^{+0.10}_{-0.16}$ \\
AS2.5$^{\dagger}$ & $1.80^{+0.43}_{-0.37}$ & $1.91^{+0.07}_{-0.07}$ & 8.97$^{+0.25}_{-0.22}$ & 10.8$^{+1.0}_{-2.5}$   \\ \hline
\end{tabular}
\\
$^{\dagger}$ Fitting is carried out in 3.5$-$8 keV because of large variation in \textit{LAXPC} data point below 3.5 keV. \\
\end{table}

\section{Testing the need for additional spectral components}
\label{appenb}
\begin{figure*}
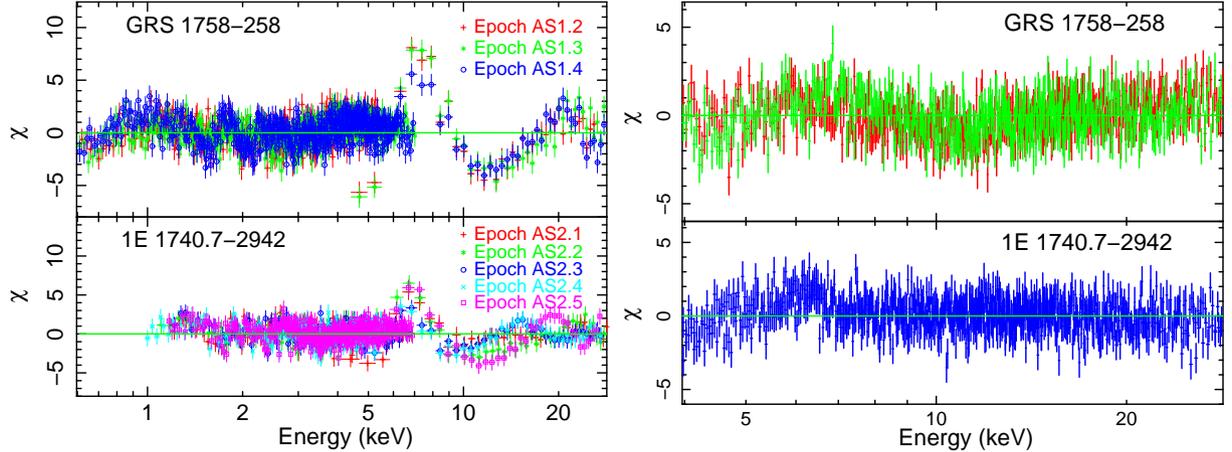

    \centering
    \includegraphics[height=8cm,angle=-90,trim={0 1.0cm 0 0},clip]{figb1a.eps}
    \includegraphics[height=8cm,angle=-90,trim={0 1.0cm 0 0},clip]{figb1b.eps}
    \caption{Residuals obtained by fitting the preliminary models without fitting for Fe line model in the \textit{AstroSat} (left) and \textit{NuSTAR} (right) spectra of both sources during different Epochs.}
    \label{appenfigb1}
\end{figure*}

\begin{figure}
        \centering
        \includegraphics[height=8cm,keepaspectratio,angle=-90,trim={0 1.0cm 0 0},clip]{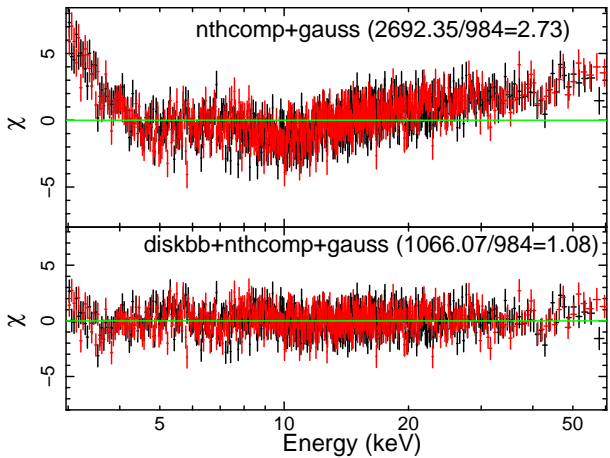}
        \caption{Residuals of the Epoch NU1.6 spectrum that is modelled using \textit{TBabs(nthComp+gauss)} in top panel and \textit{TBabs(diskbb+nthComp+gauss)} in the bottom panel. The soft excess in the residual plotted in top panel shows the need for disc component in the spectrum.}
        \label{appenfigb3}
    \end{figure}

In this section, we demonstrate the presence of additional components such as Fe emission line and thermal disk component in the Comptonization continuum of the energy spectra of both GRS 1758$-$258 and 1E 1740.7$-$2942. To validate the authenticity of the emission line, we show the improvement of the preliminary fit with the inclusion of \textit{gauss}, by comparing $\chi^{2}/dof$ of the preliminary fit with and without considering \textit{gauss} component in Table \ref{appentabb1}. The significant improvement in the $\chi^{2}/dof$ proves the need for Fe-line model in the fit. In Figure \ref{appenfigb1}, we plot the preliminary fit residual without \textit{gauss} for \textit{AstroSat} (left) and \textit{NuSTAR} (right) spectra. The stronger emission feature in all the \textit{AstroSat} spectra when compared to \textit{NuSTAR} is due to high source flux emitted by the source during these Epochs (see Table \ref{tab2} and \ref{tab3}).
\\ 
We also provide the confirmation for the presence of thermal disc component in NU1.6 spectrum. In Figure \ref{appenfigb3}, we plot the residual of the preliminary model fit without (top panel) and with (bottom panel) considering the \textit{diskbb} component. The $\chi^{2}_{red}$ improves significantly from 2.7 to 1.08, when \textit{diskbb} component is added. The presence of soft excess in the fit residual when \textit{diskbb} component is not considered, and the improvement of fit quality with the addition of \textit{diskbb} clearly indicates the need for a disc component in the lower energy. 

\begin{table}
\caption{The comparison of $\chi^{2}/dof$ ($\chi^{2}_{red}$) for the preliminary spectral fits with and without including \textit{gauss} to fit for Fe-line at $\sim6.4$ keV.} 
\begin{tabular}{ccc}
\hline
Epoch & \begin{tabular}[c]{@{}c@{}}$\chi^{2}/dof=\chi^{2}_{red}$ \\ (without Gauss)\end{tabular} & \begin{tabular}[c]{@{}c@{}}$\chi^{2}/dof= \chi^{2}_{red}$ \\ (with Gauss)\end{tabular} \\ \hline
\multicolumn{3}{c}{GRS 1758$-$258} \\ \hline
AS1.2 & $1053.24/484=2.06$                                                                         & $591.62/482=1.23$                                                                       \\
AS1.3 & $993.71/510=1.95$                                                                         & $618.13/508=1.22$                                                                       \\
AS1.4 & $925.28/535=1.73$                                                                         & $673.14/533=1.26$                                                                       \\
NU1.5 & $1972.43/1662=1.19$                                                                        & $1762.88/1659=1.06$                                                                     \\
NU1.6 & $1376.08/986=1.39$                                                                         & $1046.19/984=1.06$                                                                      \\ \hline
\multicolumn{3}{c}{1E 1740.7$-$2942} \\ \hline
AS2.1 & $178.88/144=1.24$                                                                          & $162.00/142=1.14$                                                                       \\
AS2.2 & $350.34/163=2.15$                                                                          & $168.92/161=1.05$                                                                       \\
AS2.3 & $267.37/168=1.59$                                                                          & $166.61/166=1.00$                                                                       \\
AS2.4 & $331.48/259=1.28$                                                                          & $254.68/257=0.99$                                                                       \\
AS2.5 & $576.62/342=1.69$                                                                          & $328.29/340=0.97$                                                                       \\
NU2.6 & $1050.35/845=1.24$                                                                         & $896.34/842=1.06$                                                                       \\ \hline
\end{tabular}
\label{appentabb1}
\end{table}

\section{Non-relativistic Reflection Modeling} 
\label{appen}
 \begin{figure}
        \centering
        \includegraphics[height=8cm,keepaspectratio,angle=-90,trim={0 2.5cm 0 0},clip]{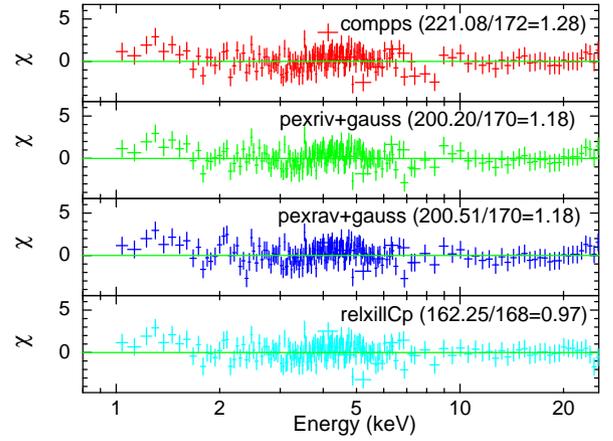}
        \caption{Residual of the different reflection model fits, fitted to the \textit{AstroSat} spectrum in 0.8$-$30 keV belonging to Epoch AS2.3 observation. $\chi^{2}_{red}$ ($\chi^{2}/dof$) value for each of these fit is mentioned in the plot.}
        \label{appenfigc1}
    \end{figure}
    \begin{table*}
\caption{Model parameters of GRS 1758$-$258 obtained by fitting Comptonization and non-relativistic reflection models to the Epoch NU1.5 spectrum in 3$-$79 keV energy range. In all of these fits, $N_{H}$ value is frozen to $2.5\times10^{22}$ atoms cm$^{-2}$. Errors of all the parameters are obtained with 90$\%$ confidence.}
\begin{tabular}{ccccc}
\hline
Parameter                  & \texttt{nthComp+gauss}    & \texttt{pexriv+gauss}          & \texttt{pexrav+gauss}   & \texttt{compPS}                     \\ \hline
$\Gamma$               & $1.64\pm0.01$    & $1.65\pm0.01$         & $1.66\pm0.01$  & $-$                               \\
$kT_{e}$ (keV)             & $48.2^{+11.1}_{-6.0}$ & $-$                   & $-$            & $54.5^{+6.6}_{-0.7}$  \\
lineE (keV)                & $6.11\pm0.19$    & $6.00^{+0.26}_{-0.25}$ & $6.83\pm0.24$  & $-$                                          \\
$\tau$                     & $-$                & $-$                   & $-$              & $2.99\pm0.02$                              \\
$rel\_refl$ ($\Omega/2\pi$) & $-$                & $0.19\pm0.03$        & $0.25\pm0.05$  & $0.17\pm0.02$                     \\
$\chi^{2}/dof$             & $1875.40/1659$     & $1764.53/1658$        & $1697.43/1660$ & $1813.76/1661$              \\ \hline
\end{tabular}
\label{tabc1}\\
\raggedright
\hspace{3cm}
\end{table*}
\begin{table*}
\caption{Spectral fit parameter of the source 1E 1740.7$-$2942 during Epoch AS2.3. The \textit{AstroSat} energy spectrum is modelled with different Comptonization and non-relativistic reflection models in 0.6$-$30 keV energy band. Errors of the parameters are quoted with 90$\%$ confidence.}
\begin{tabular}{ccccc}
\hline
                                                                             Parameter & \texttt{nthComp+gauss}          & \texttt{pexriv+gauss}           & \texttt{pexrav+gauss}            & \texttt{compPS}                      \\ \hline
\begin{tabular}[c]{@{}c@{}}$N_{H} \times 10^{22}$\\  (atoms cm$^{-2}$)\end{tabular} & $11.0\pm0.2$ & $10.1^{+0.6}_{-0.5}$ & $11.4\pm0.6$ & $11.6\pm0.3$      \\
$\Gamma$                                                                  & $1.97^{+0.02}_{-0.01}$          & $2.21\pm0.01$          & $2.32\pm0.01$           & $-$                                            \\
$kT_{e}$ (keV)                                                                & $4.95^{+0.20}_{-0.21}$       & $-$                    & $-$                     & $35.6^{+10.5}_{-7.0}$          \\
lineE (keV)                                                                   & $6.00\pm0.05$          & $6.77\pm0.11$  & $6.82\pm0.15$           & $-$                                                \\
$\tau$                                                                        & $-$                      & $-$                    & $-$                       & $0.45^{+0.33}_{-0.09}$                                    \\
$rel\_refl$ ($\Omega/2\pi$)                                                    & $-$                      & $0.63^{+0.48}_{-0.28}$ & $0.93^{+0.32}_{-0.28}$  & $0.60^{+0.15}_{-0.16}$                        \\
$\chi^{2}/dof$                                                                & $202.09/200$           & $200.20/170$           & $200.51/170$             & $221.08/172$                             \\ \hline
\end{tabular}
\\
\hspace{3cm}
\raggedright
$^{f}$ Frozen parameter
\label{tabc2}
\end{table*}
This section explains the details of modeling carried out using different non-relativistic reflection models. Initially, we model the spectra of both source using the non-relativistic reflection models such as \textit{pexriv} and \textit{pexrav} \citep{1995MNRAS.273..837M}. While \textit{pexrav} accounts for reflection in the spectrum by assuming a neutral-disc ($\xi=0$), \textit{pexriv} considers disc to be ionized ($\xi$ is free parameter). Here, we show the results obtained by fitting these models to Epoch NU1.5 spectrum of GRS 1758$-$258 (Table \ref{tabc1}) and AS2.3 spectrum of 1E 1740.7$-$2942 (Table \ref{tabc2}). During these fits, we fix the cosine of inclination angle to 0.6. Fe abundance of GRS 1758$-$258 is assumed to be the same as that of solar abundance, whereas that of 1E 1740.7$-$2942 is 2 times solar metal abundance. As we could not constrain the $\xi$ parameter from \textit{pexriv}, we fixed its value to 1000 erg cm s$^{-1}$ assuming a partially ionized disc. Since these two models do not include line emission, we added \textit{gauss} at $\sim6.4$ keV. \textit{pexriv} and \textit{pexrav} resulted in $R_{f}$ of $0.19\pm0.03$ and $0.25\pm0.05$ respectively in case of GRS 1758$-$258 and $0.63^{+0.48}_{-0.28}$ and $0.93^{+0.32}_{-0.28}$ in case of 1E 1740.7$-$2942. We then model the same spectrum using \textit{compPS} model \citep{1996ApJ...470..249P} that computes the Comptonization spectra for different geometries which also accounts for reflection. The reflected radiation from the cool disc is quantified by the parameter $rel\_refl$ in the units of $2 \pi$. Here, the seed photons are assumed to be in the form of a disc blackbody with the optical depth $\tau$ as a free parameter. In all our fits, we consider a Maxwellian distribution of electrons in plasma with a spherical geometry. The seed photon temperature $kT_{bb}$ is fixed to 0.1 keV. The disc is assumed to be partially ionized ( $\xi=1000$ erg cm s$^{-1}$) and to have an inclination angle of $60^{\circ}$ in both sources. This model provides constraint on $kT_{e}$ of GRS 1758$-$258 to be $54^{+7}_{-1}$ keV and that of 1E 1740.7$-$2942 to be $36^{+10}_{-7}$ keV. Optical depth ($\tau$) of GRS 1758$-$258 and 1E 1740.7$-$2942 are constrained to be $2.99\pm0.02$ and $0.45^{+0.33}_{-0.09}$ respectively. \\
Even though the non-relativistic models provide a good fit, we employ \textit{relxillCP} model to describe the reflection properties of both BHs because of its large spectral flexibility that allows us to constrain physical parameters. The relativistic reflection model is preferred also because of the inclusion of fundamental physical processes that are generally absent in non-relativistic models. In order to have a statistical comparison of different reflection models, we compare the residuals and the reduced chi-square ($\chi^{2}_{red}$) of the spectral fits using these models for all the observations. From this, we find that \textit{pexiv/pexrav+gauss} model yields a better fit in few observations (Example: Epoch NU1.5, Table \ref{tabc1}) while in others, best-fit is obtained by \textit{relxillCp}. As an example, we plot the residual of different models along with the $\chi^{2}/dof=\chi^{2}_{red}$ for Epoch AS2.3 observation in Figure \ref{appenfigc1} where we see that \textit{relxillCp} yields better fit when compared to all other models in this observation.


\bsp	
\label{lastpage}
\end{document}